\documentclass[twocolumn,showpacs,graphicx,floatfix,eqsecnum,prb,twocolumn,hyperref=pdftex,amsmath,longbibliography]{revtex4-1}
\usepackage{graphics}
\usepackage{showlabels}
\usepackage{amsmath}
\usepackage{amsfonts}
\usepackage{bm}
\usepackage{color}
\usepackage{bbm}
\usepackage{amssymb}
\usepackage{array}
\usepackage{graphicx}
\usepackage{graphics}
\usepackage{dcolumn}
\usepackage{bm}\usepackage{varioref}
\usepackage{hyperref}
\usepackage{comment}
\hypersetup{colorlinks=true,linkcolor=blue,anchorcolor=blue,citecolor=blue,filecolor=blue,urlcolor=blue,bookmarksnumbered=true,pdfview=FitB}
\newcommand{\nn}{\nonumber}
\newcommand{\bk}{{\bf k}}
\newcommand{\bp}{{\bf p}}
\newcommand{\bq}{{\bf q}}

\newcommand{\bl}{{\bf l}}
\newcommand{\vare}{\varepsilon}
\newcommand{\beq}{\begin{equation}}
\newcommand{\eeq}{\end{equation}}
\newcommand{\bea}{\begin{eqnarray}}
\newcommand{\eea}{\end{eqnarray}}
\newcommand{\bv}{{\bf v}}
\newcommand{\br}{{\bf r}}

\newcommand{\bwt}{\begin{widetext}}
\newcommand{\ewt}{\end{widetext}}
\begin{document}
\title{Lorentz ratio of a compensated metal}
\author{Songci Li$^{1,2}$ and Dmitrii L. Maslov$^2$}
\affiliation{$^1$National High Magnetic Field Laboratory, Tallahassee, Florida 32310, USA\\
$^2$Department of Physics, University of Florida, P. O. Box 118440, Gainesville, Florida 32611, USA}
\begin{abstract}
A violation of the Wiedemann-Franz law in a metal can be quantified by comparing the Lorentz ratio, $L=\kappa\rho/T$, where $\kappa$ is the thermal conductivity and $\rho$ is the electrical resistivity, with the universal Sommerfeld constant, $L_0=(\pi^2/3) (k_B/e)^2$. 
We obtain the Lorentz ratio of a clean compensated metal with intercarrier interaction as the dominant scattering mechanism  by solving exactly the system of coupled integral Boltzmann equations. The Lorentz ratio is shown to assume a particular simple form in the forward-scattering limit: $L/L_0=\overline{\Theta^2}/2$, where $\Theta$ is the scattering angle.  In this limit, $L/L_0$ can be arbitrarily small.  We also show how the same result can be obtained without the benefit of an exact solution. We discuss how a strong downward violation of the Wiedemann-Franz law in a type-II Weyl semimetal WP$_2$ can be explained within our model.
\end{abstract}
\date{\today}
\maketitle
\section{Introduction}
\label{sec:intro}
According to the Wiedemann-Franz law (WFL),\cite{Ziman:EP,abrikosov:book,physkin} the Lorentz ratio $L(T)=\kappa\rho/ T$, where $\kappa$ is the thermal conductivity and $\rho$ is the electrical resistivity of a metal, is given by the universal Sommerfeld 
constant  \bea
L_0=(\pi^2/3)(k_B/e)^2.\label{L0}
\eea  The WFL holds if electron scattering is elastic,\cite{Ziman:EP,physkin} such that the relaxation times of the charge current, $\tau_\rho$, and of the thermal current, $\tau_\kappa$, are the same. The WFL  holds 
both at very low temperatures, when electrons are scattered mostly by disorder, and at temperatures above the Debye one,
% $T\gg \Theta_D$ ($\Theta_D$ is the Debye temp), 
when scattering of electrons by phonons becomes quasielastic.\footnote {Historically, the WFL was observed first in the regime of quasielastic electron-phonon scattering\cite{WF:1853} because cryogenic technologies were not available in 1853} At intermediate temperatures, %$T\ll\Theta_D$, 
%the 
%inelastic 
scattering is inelastic,
%renders 
%a discrepancy between 
the two relaxation times differ from each other, and the WFL is violated.

In the case of inelastic electron-phonon scattering, the difference between $\tau_\rho$ and $\tau_\kappa$ is due to the fact that electrons are scattered 
%scattering with 
by acoustic phonons 
%whose 
with group velocity $s$ and typical momenta $q\sim T/s$,
%\ll p_F$ 
%, 
the latter being much smaller than the
Fermi momentum, $p_F$. Therefore, $\tau_\rho$ is longer than the single-particle relaxation time, $\tau\propto T^{-3}$, due to the $1-\cos\Theta$ factor, which filters out small-angle scattering events, and 
%the typical momenta of electrons. 
%Hence, 
$\rho\propto\tau^{-1}_\rho\propto T^5$. On the other hand, since every collision is effective in energy relaxation, we have $\tau_\kappa\sim\tau$, and the thermal resistivity $w\equiv T/\kappa$
scales as $\tau^{-1}
\propto T^3$.  As a result, 
one obtains a downward violation of the WFL, i.e., 
 $L(T)<L_0$, which is 
often 
observed in elemental 
%and 
%heavy-fermion 
metals.\cite{white:1960,zhang:2000,yao:2017}

Downward deviations from the WFL are also observed in cases when the electron-electron interaction is known (or suspected) to be the dominant scattering mechanism.
For example, the values of $L(T)<L_0$ were measured in the normal state of the cuprate superconductors,\cite{zhang:2000}
%YBa$_2$Cu$_3$O$_6.95$, 
in heavy-fermion metals,\cite{lussier:1994,paglione:2005,tanatar:2007,dong:2013,machida:2013}
% CeRhIn$_5$ (Ref.~\onlinecite{paglione:2005}), 
near a magnetic-field-tuned quantum critical end point,\cite{ronning:2006}
% in Sr$_3$Ru$_2$O$_7$ (Ref.~\onlinecite{ronning:2006}), 
and in a candidate type-II Weyl semimetal WP$_2$ (Refs.~\onlinecite{Gooth:2017} and \onlinecite{Jaoui:2018}). 
\footnote{A drastic upward violation of the WFL is observed in graphene at the charge neutrality point\cite{crossno:2016} and is understood as arising from the weakness of electron-hole  generation-recombination processes, which control the thermal conductivity of graphene \cite{mueller:2008b,foster:2009,crossno:2016}}
The interpretation of such experiments is complicated by the fact the charge current can be degraded only by umklapp or interband scattering, whereas the thermal current is degraded already by intraband normal scattering, but is affected by umklapp and interband scattering as well. Consequently, the Lorentz ratio depends on the ratio of the umklapp and normal scattering rates which, in turn, is very sensitive to the geometry and topology of the Fermi surface (FS) and thus highly non-universal. However, if umklapp scattering is excluded because, e.g., the FS is too small\cite{abrikosov:book} or the interaction is of a long range,\cite{maslov:2011,pal:2012b,maslov:2017b} the situation is somewhat simplified because normal scattering in a metal with anisotropic FS affects both electrical and thermal currents. In general, however, one still needs to introduce momentum-relaxing scattering, e.g., by impurities or phonons, which ultimately renders the electrical conductivity finite.

%In this case, the downward violation of the WFL indicates small-angle scattering, similarly to the electron-phonon case described above. Unlike electron-phonon scattering, however, small-angle {\em e-e} scattering occurs not because of the phase-space constraints but only if the {\em e-e} interaction is long-ranged. 

There is one but very important exception to this rule, namely a compensated metal (CM) with equal numbers of electrons and holes. 
The electrical conductivity of a CM is rendered finite already by normal scattering between electrons and holes (the Baber mechanism \cite{baber:1937}), while its thermal conductivity contains contributions from both intra- and interband scattering processes. At high enough temperatures, the electron-hole and electron-electron interactions control both electrical and thermal transport without the help of additional momentum-relaxing processes, and one can make certain statements about the magnitude of the Lorentz ratio within a tractable model. 

In this paper, we calculate the Lorentz ratio of a CM, assuming that the intercarrier interaction is the dominant scattering mechanism. Our particular goal is to understand 
recent observation of an abnormally small ($\approx  0.2L_0$) Lorentz ratio in bulk WP$_2$ (Ref.~\onlinecite{Jaoui:2018}).  We will argue that this can be attributed to weak screening in this material.
In a broader context, the family of CMs is quite large:\cite{fawcett:1963} it includes many metals and semimetals with an even number of electrons per unit cell, e.g., Mg, Zn, Cd, Bi, graphite, etc.  A relatively recent addition to the family are iron-based superconductors in their parent states,\cite{stewart:2011} most of which have compensated electron and hole pockets. 
 Finally, the most recently discovered members of the family are type-II Weyl semimetals (Ref.~\onlinecite{soluyanov:2015}), e.g., WP$_2$ (Ref.~\onlinecite{WP2:compensation,comment:compensated}). The interest in electron transport in CMs has been rekindled by recent observations of extremely large magnetoresistance \cite{kumar:2017b,fauque:2018} and possible realization of the hydrodynamic (Gurzhi\cite{Gurzhi:1968}) flow regime\cite{Gooth:2017,Jaoui:2018,coulter:2018} in these materials. Having even a simple model for electrical and thermal transport in CMs would be useful for understanding the unique properties of CMs.

In what follows, the electron band ($1$) and hole band ($2$) will be assumed to have parabolic dispersions, 
\bea
  \varepsilon_{1,\mathbf{p}}=\frac{\left(\mathbf{p}-\bp_0/2\right)^2}{2m_1},\, \varepsilon_{2,\mathbf{p}}=-\frac{\left(\mathbf{p}+\bp_0/2\right)^2}{2m_2}+\Delta, \label{bands}
\eea
where $\Delta$ is the energy offset, and $m_{1(2)}$ is the electron (hole) effective mass.  In a CM, the electron and hole density are equal, $n_1=n_2=n$, hence the Fermi momentum and Fermi energy are given by $p_F=(3\pi^2 n)^{1/3}$ and $\varepsilon_F=\Delta m_2/(m_1+m_2)$,
correspondingly. We will only be interested in the degenerate regime of $T\ll \varepsilon_F$.  (Throughout the paper we take $\hbar=k_B=1$, unless  specified otherwise.) The separation between the electron and hole bands ($p_0$) is assumed to be much larger than the (inverse) radius of the interaction, so interband transfer of carriers is not allowed.

The rest of the paper is organized as follows. In Sec.~\ref{sec:1} we briefly introduce the Boltzmann equation (BE). 
In Sec.~\ref{sec:2} we show that a system of BEs allows for a simple solution for the case of forward scattering and obtain the corresponding results for the electrical and thermal conductivities. 
In Sec.~\ref{sec:3} we find the exact results for the electrical and thermal conductivities, using the method of Refs.~\onlinecite{abrikosov:1959,Sykes,Wilkins,Sykes_2} for an arbitrary interaction potential, and compare these results to the approximate ones obtained in Sec.~\ref{sec:2}.  In Sec.~\ref{sec:lorentz} we discuss the Lorentz ratio for a number of model and make a connection to an experiment on WP$_2$ (Ref.~\onlinecite{Jaoui:2018}). Our conclusions are given in Sec.~\ref{concl}.
\section{Linearized Boltzmann Equation}\label{sec:1}
To set the stage, we briefly introduce the BE for a single band case. The generalization to a two band case is straightforward. The semiclassical BE for the distribution function $f_\bp(\br,t)$ is written as
\begin{eqnarray}
	\frac{\partial f_{\mathbf{p}}}{\partial t}+\frac{\partial \varepsilon_{\mathbf{p}}}{\partial\mathbf{p}}\cdot\frac{\partial f_{\mathbf{p}}}{\partial \mathbf{r}}-\frac{\partial \varepsilon_\bp}{\partial\mathbf{r}}\cdot\frac{\partial f_{\mathbf{p}}}{\partial \mathbf{p}}=-\mathcal{I}[f_{\mathbf{p}}], \label{eq:BE_general}
\end{eqnarray}
where $\mathcal{I}$ is the collision integral which accounts for scattering processes. 

If Eq.~(\ref{eq:BE_general}) describes a Fermi liquid (FL), $\vare_\bp$ on its left-hand side is to be understood as the non-equilibrium quasi-particle energy, which is related to $f_\bp$ via the self-consistent equation of the FL theory.\cite{lifshitz:1980}  As a result,  the left-hand side of linearized Eq.~(\ref{eq:BE_general}) contains two corrections to the equilibrium distribution function.\cite{nozieres,physkin} The ``bare'' one, $\delta n_\bp$, is defined by writing $f_\bp$ as $f_\bp=n_\bp+\delta n_\bp$, where $n_\bp\equiv n_F\left(\vare_\bp^{(0)}\right)$ is the equilibrium Fermi function and $\vare_\bp^{(0)}$ is the equilibrium quasiparticle energy. The time derivative on the left-hand side of linearized Eq.~(\ref{eq:BE_general}) contains $\delta n_\bp$. The ``renormalized'' one,  $\delta\bar n_\bp$, is related to the bare one via  
\bea
\delta\bar n_\bp=\delta n_\bp-\frac{\partial n_\bp}{\partial \vare_\bp}\int_{\bp'} F^{s}(\bp,\bp')\delta n_{\bp'},
\eea
where $\int_\bp$ is a shorthand for $\int d^Dp/(2\pi)^D$ and $F^{s}(\bp,\bp')$ is the spin-symmetric part of the Landau interaction function.
 On the other hand, the gradient term in linearized  Eq.~(\ref{eq:BE_general}) and macroscopic observables contain
 $\delta\bar n_\bp$. For example the charge current is given by ${\bf j}=e\int_\bp\bv_\bp\delta\bar n_\bp$. However, the collision integral can also be expressed via $\delta\bar n_\bp$ (Refs.~\onlinecite{nozieres,physkin}). Therefore, if the time dependence can be ignored, $\delta n_\bp$ does not appear in the theory,  while $\delta\bar n_\bp$ plays the role of a proper distribution function. As we will be interested only in {\em dc} transport, $\delta n_\bp$ in the remainder of the paper is to be understood as $\delta\bar n_\bp$, with bar suppressed for brevity. In this way, the kinetic equation for a FL coincides with that for the Fermi gas, the only difference being that $\bv_\bp=\partial_\bp\vare_\bp$ in this equation is to be understood as the renormalized Fermi velocity.
 
The collision integral describing electron-electron interaction in a single-band metal can be written as 
\begin{eqnarray}
	&&\mathcal{I}[f_{\mathbf{p}}] = \int_{\mathbf{k}}\int_{\mathbf{p}'}\int_{\mathbf{k}'}W_{\mathbf{p}\mathbf{k\rightarrow}\mathbf{p}'\mathbf{k}'}\delta(\varepsilon_{\mathbf{p}}+\varepsilon_{\mathbf{k}}-\varepsilon_{\mathbf{p}'}-\varepsilon_{\mathbf{k}'}) \nonumber\\
	&&\times\delta(\mathbf{p}+\mathbf{k}-\mathbf{p}'-\mathbf{k}')\Big[f_{\mathbf{p}}f_{\mathbf{k}}(1-f_{\mathbf{p}'})(1-f_{\mathbf{k}'}) \nonumber \\
	&& -f_{\mathbf{p}'}f_{\mathbf{k}'}(1-f_{\mathbf{p}})(1-f_{\mathbf{k}})\Big],
\end{eqnarray} 
where $W_{\mathbf{p}\mathbf{k\rightarrow}\mathbf{p}'\mathbf{k}'}$ is the scattering probability of  intercarrier scattering. 
%One can write $f_{\mathbf{p}}$ as $f_{\mathbf{p}}=n_{\mathbf{p}}+\delta n_{\mathbf{p}}$, where $n_{\mathbf{p}}$ is the equilibrium distribution function and $\delta n_{\mathbf{p}}$ is the non-equilibrium correction. 
The collision integral can be linearized by defining 
\begin{eqnarray}
	\delta n_{\mathbf{p}}\equiv -T\frac{\partial n_\bp}{\partial\varepsilon_{\mathbf{p}}}g_{\mathbf{p}}=n_{\mathbf{p}}(1-n_{\mathbf{p}})g_{\mathbf{p}}, \label{g}
\end{eqnarray}
which yields\cite{abrikosov:book}
\begin{eqnarray}
	&&\mathcal{I}[g_{\mathbf{p}}] = \int_{\mathbf{k}}\int_{\mathbf{p}'}\int_{\mathbf{k}'}W_{\mathbf{p}\mathbf{k\rightarrow}\mathbf{p}'\mathbf{k}'}\nonumber \\
	&&\times n_{\mathbf{p}}n_{\mathbf{k}}(1-n_{\mathbf{p}'})(1-n_{\mathbf{k}'})(g_{\mathbf{p}}+g_{\mathbf{k}}-g_{\mathbf{p}'}-g_{\mathbf{k}'}) \nonumber \\
	&&\times\delta(\varepsilon_{\mathbf{p}}+\varepsilon_{\mathbf{k}}-\varepsilon_{\mathbf{p}'}-\varepsilon_{\mathbf{k}'})\delta(\mathbf{p}+\mathbf{k}-\mathbf{p}'-\mathbf{k}').
\end{eqnarray} 

\section{Transport coefficients in the forward-scattering limit}\label{sec:2}
In this section we examine the forward-scattering limit which is relevant, e.g., for the case of a weakly screened Coulomb interaction, or
to scattering by ferromagnetic or nematic fluctuations near a corresponding quantum phase transition.\cite{maslov:2011} 
\subsection{Electrical conductivity}
\label{sec:rho_for}
In the presence of an external electric field, the two coupled BEs for the electron and hole bands can be readily obtained as a generalization of Eq.~(\ref{eq:BE_general}),
\begin{subequations}
	\begin{eqnarray}
	-e\mathbf{E}\cdot\mathbf{v}_{1,\bp}\frac{\partial{n}_F}{\partial\vare_{1,\bp}} &=& -\mathcal{I}^{12}[g_1,g_2], \label{eq:BE_rho_1}\\
	-e\mathbf{E}\cdot\mathbf{v}_{2,\bk}\frac{\partial{n}_F}{\partial\vare_{2,\bk}} &=& -\mathcal{I}^{21}[g_1,g_2],
	\label{eq:BE_rho_2}
\end{eqnarray}
\end{subequations}
where $\bv_{j,{\bf l}}$ is the group velocity of the $j\text{th}$ band and $g_{1,2}$ are defined as in Eq.~(\ref{g}) for each of the two bands.
 The collision integrals $\mathcal{I}^{12}$ and $\mathcal{I}^{21}$ describe the Baber-type\cite{baber:1937} interband scattering between electron and holes: \begin{widetext}
	\begin{eqnarray}
	\mathcal{I}^{12}[g_1,g_2] = \int_{\bk}\int_{\bp'}\int_{\bk'}&&W^{12}_{\bp\bk\to\bp'\bk'}
	n_{1,\bp}n_{2,\bk}(1-n_{1,\bp'})(1-n_{2,\bk'})
	\left[g_{1,\bp}+g_{2,\bk}-g_{1,\bp'}-g_{2,\bk'}\right]
	\delta(\varepsilon_{1,\bp}+\varepsilon_{2,\bk}-\varepsilon_{1,\bp'}-\varepsilon_{2,\bk'}) \nonumber \\
	&&\times\delta(\bp+\bk-\bp'-\bk'),\label{I12}
\end{eqnarray}
\end{widetext}
and $\mathcal{I}^{21}$ is obtained from $\mathcal{I}^{12}$ by interchanging the band indices.
In our model of parabolic bands [cf.~Eq.~(\ref{bands})], intraband scattering does not affect the electrical conductivity, and the corresponding collision integrals  have been dropped. 

It is convenient to introduce the momentum transfer $\mathbf{q}$ and energy exchange $\omega$, such that $\bp'=\bp-\bq$, $\bk'=\bk+\bq$,
$\vare_{j,\bp'}=\vare_{j,\bp}-\omega$, and $\vare_{j,\bk'}=\vare_{j,\bk}+\omega$, where $j=1,2$. In the FL regime, the scattering probability can be taken as independent of $\omega$. If electrons interact via a potential $V(\bq)$, the symmetrized scattering probability  for a carrier with spin $\alpha$ is given by
\bea
%&&W^{11}_{\bp\bk\to\bp'\bk'}\equiv W_{11}(\bq);\,W^{22}_{\bp\bk\to\bp'\bk'}\equiv W_{22}(\bq)\;\nonumber\\
&&W^{12}_{\bp\bk\to\bp'\bk'}=W^{21}_{\bp\bk\to\bp'\bk'}\nn\\
&&=2\pi \sum_{\beta\gamma\delta}\left\vert V(\bq)\delta_{\alpha\gamma}\delta_{\beta\delta}-V(\bp-\bk-\bq)\delta_{\alpha\delta}\delta_{\beta\gamma}\right\vert ^2.\label{sca}
\eea
For a long-range interaction, the second (exchange) term under $|\dots|$ in the equation above can be neglected, in which case $W^{12}$ depends only
on momentum transfer $\bq$ but not on the initial momenta $\bp$ and $\bk$. In addition, if the system is isotropic, $V(\bq)=V(q)$ and \bea
W^{12}_{\bp\bk\to\bp'\bk'}=W^{21}_{\bp\bk\to\bp'\bk'}\equiv W(q)=4\pi|V(q)|^2.\label{W_q}
\eea

 After these steps, the collision integral can be rewritten as
\begin{widetext}
\begin{eqnarray}
		\mathcal{I}^{12}[g_1,g_2] &=& \int_{\bk}\int_{\mathbf{q}}\int d\omega W^{12}(\mathbf{q})n_{1,\bp}n_{2,\bk}(1-n_{1,\bp-\bq})(1-n_{2,\bk+\bq})
	\left(g_{1,\bp}+g_{2,\bk}-g_{1,\bp-\bq}-g_{2,\bk+\bq}\right) \nn \\ 
	&&\times\delta(\varepsilon_{1,\bp}-\varepsilon_{1,\bp-\bq})\delta(\varepsilon_{2,\bk}-\varepsilon_{2,\bk+\bq}), \label{eq:collb}
\end{eqnarray}
\end{widetext}
where it is understood that $n_{j,{\bf l}\pm \bq}=n_F(\vare_{j,{\bf l}}\pm\omega)$.
 In the equation above, we have also neglected $\omega$ in the arguments of the $\delta$-functions which ensure energy conservation. The reason is that the scaling dimensions of the two energy integrals (over $\vare_{\bk}$ and $\omega$) already give the expected $T^2$ scaling of the collision integral; keeping $\omega$ in other places would give only subleading terms. 

The nonequilibrium part of the distribution function can be parameterized as 
\begin{eqnarray}
g_{j,\bl}=-\frac{e}{T}(\mathbf{v}_{j,\bl}\cdot\mathbf{E})\,\varphi_j\left(\frac{\xi_{j,\bl}}{T}\right), \label{eq:psi_rho}
\end{eqnarray} 
where $j=1,2$, $\xi_{j,\mathbf{l}}\equiv \varepsilon_{j,\mathbf{l}}-\vare_F$ and $\varphi_{j}(x)$ is an even function of its argument. 
In general, one needs to solve the system of integral equations for $\varphi_j(x)$, which is what we will do in Sec.~\ref{sec:3}. In the forward-scattering limit, however, the procedure can be  simplified because in this case the energy relaxation is much faster than the momentum one: a thermally excited carrier first descends to the FS and then diffuses around the FS via small-angle scattering events. As a result, the nonequilibrium  part of the distribution function depends primarily on the direction of the momentum, while the dependence on its magnitude (energy) is much weaker and can to be taken into account only to leading order that ensures the symmetry requirements. The simplest choice for $\varphi_j(x)$ is just a constant:
\begin{equation}
	\varphi_j\left(x\right)=\text{const}\equiv a_j.
\end{equation}
(The same argument was used in Refs.~\onlinecite{maslov:2011,pal:2012b} to find the conductivity of an uncompensated two-band system.)

 Substituting $g_j$ in Eqs.~(\ref{eq:BE_rho_1}) and (\ref{eq:BE_rho_2}), 
 we obtain a single equation relating $a_1$ and $a_2$: 
\begin{eqnarray}
	\frac{a_1}{m_1}+\frac{a_2}{m_2}=\frac{48\pi^2 \hbar^6 p^3_F}{m^2_1 m^2_2T^2}\frac{1}{\int dq q^2W^{12}(q)}. \label{eq:c_rho_1c_rho_2}
\end{eqnarray}
Although Eq.~(\ref{eq:c_rho_1c_rho_2}) does not allow one to find $a_1$ and $a_2$ independently, it suffices to determine the total electric current density, which is proportional to the same combination $a_1/m_1+a_2/m_2$:
\begin{eqnarray}
	\mathbf{j} &=& 2\sum_j \int\frac{d^3 p_j}{(2\pi)^3}(-e)\mathbf{v}_j(-Tn^{\prime}_j g_j), \nonumber\\
	&=& ne^2\left(\frac{a_1}{m_1}+\frac{a_2}{m_2}\right)\mathbf{E}. \label{eq:current}
\end{eqnarray}
Using Eqs.~(\ref{eq:c_rho_1c_rho_2}) and (\ref{eq:current}), we obtain the electrical resistivity as
\begin{eqnarray}
	\rho=\frac{1}{ne^2}\frac{m^2_1 m^2_2T^2}{48\pi^2p^3_F}\int dqq^2W(q). \label{eq:rho_forward}
\end{eqnarray}
A factor of $q^2$ in the integrand is the familiar ``transport factor'' that filters out small-angle scattering events. In Sec.~\ref{sec:3} we will show that the exact result for $\rho$ is indeed reduced to Eq.~(\ref{eq:rho_forward})  in the forward-scattering limit.

A screened Coulomb interaction is described by the potential
\bea
V(q)=\frac{4 \pi e^2}{\epsilon_0}\frac{1}{q^2+\varkappa^2},\label{Vq}
\eea
where $\varkappa$ is the inverse screening length and $\epsilon_0$ is the dielectric constant at zero frequency. In this case, 
\begin{eqnarray}
	\rho=\frac{\pi^2}{3}\frac{1}{ne^2}\frac{T^2 m^2_1m^2_2e^4}{
	\epsilon_0^2 p^3_F \varkappa}.\label{eq:rho_forward_1}
\end{eqnarray}
The forward-scattering approximation is justified for $\varkappa\ll p_F$, which is also a condition
for writing $V(q)$ as in Eq.~(\ref{Vq}). Note that if $m_1\neq m_2$ the resistivity cannot be cast into a Drude form, i.e., $\rho=m/ne^2\tau$, because $m$ and $\tau$ cannot be defined uniquely.

In 2D, the corresponding average of the scattering probability 
\begin{eqnarray}
	\int dq q \frac{W(q)}{1-(q/2p_F)^2} \label{eq:2pF}
\end{eqnarray}
diverges logarithmically at $q=2p_F$. The denominator in Eq.~(\ref{eq:2pF}) is obtained as a result of the angular integration of the two delta-functions in Eq.~(\ref{eq:collb}):
\begin{eqnarray}
&& \int d\Omega_{\mathbf{p}\mathbf{q}} \delta(\varepsilon_{1,\bp}-\varepsilon_{1,\bp-\bq})\int d\Omega_{\mathbf{k}\mathbf{q}}\delta(\varepsilon_{2,\bk}-\varepsilon_{2,\bk+\bq}) \nonumber \\
&& \propto\frac{1}{q^2}\frac{1}{1-(q/2p_F)^2}.
\end{eqnarray}
Cutting off the divergence at $|q-2p_F|\sim T/v_F$, we obtain $\rho\propto T^2\ln T$.
However,  the logarithmic factor is just the first term in the series for the Cooper scattering amplitude in the backscattering channel with $(\bp,-\bp\to-\bp,\bp)$ (Ref.~\onlinecite{shekhter:2006,chubukov:2007}). Resumming this series, one obtains  
\bea
\rho\propto \frac{T^2}{\ln^2 T}.\label{rho2D}
\eea
This result is the same as that for the (inverse) shear viscosity of a single-band 2D FL.\cite{novikov:2006}
Strictly speaking, the forward-scattering approximation is not valid for large momentum transfers ($q\approx 2p_F$), but the scaling form in Eq.~(\ref{rho2D}) remains correct even beyond this approximation. 

Note that the $2p_F$ singularity in Eq.~(\ref{eq:2pF}) comes about as a product of two square-root singularities: $1/\sqrt{1-(q/2p_{F,1})^2}$ and 
$1/\sqrt{1-(q/2p_{F,2})^2}$ with $p_{F,1}=p_{F,2}=p_F$. If a metal is not compensated, i.e., $p_{F,1}\neq p_{F,2}$, each of the square-root singularities is integrable on its own and there is no logarithmic factor in the result. In this case, however, one needs to introduce a momentum-relaxing process, e.g., impurity scattering, to render the resistivity finite. As a result, the resistivity increases with temperature from its residual value  at the lowest temperatures towards another impurity-controlled limiting value at the highest temperatures.\cite{maslov:2011,pal:2012b} If the band masses differ substantially, so do the low- and high-temperature limits of the resistivity, and there is a well-defined intermediate region in which $\rho$ scales just as $T^2$ even in 2D, without an extra logarithmic factor.  Also, if a 2D metal is compensated but has an unequal number of electron and hole pockets (as it is the case, e.g., for the parent state of iron-based superconductors\cite{stewart:2011}), the Fermi momenta of electrons and holes are different and, as result, the resistivity also scales just as $T^2$, without an extra logarithmic factor.

\subsection{Thermal Conductivity}
\label{sec:kappa_for}
The driving term for thermal transport is 
\bea
-\frac{\partial n}{\partial \varepsilon_{\mathbf{p}}}\mathbf{v}_\mathbf{p}\cdot\boldsymbol{\nabla}T\,\frac{\xi_{\mathbf{p}}}{T},\,\xi_{\mathbf{p}}=\varepsilon_{\mathbf{p}}-\varepsilon_F.
\eea 
The relevant scattering processes  in a two-band system include both intra- and interband scattering. 
Consequently, the linearized BEs for the two-band system read
\begin{subequations}
	\begin{eqnarray}
	-\frac{\partial n_{1,\bp}}{\partial \xi_{1,\bp}}\mathbf{v}_{1,\bp}\cdot\boldsymbol{\nabla}T\,\frac{\xi_{1,\bp}}{T} &=& -\mathcal{I}^{11}[g_1]-\mathcal{I}^{12}[g_1,g_2],\nn\\ \label{eq:BE_thermal_1}\\
    -\frac{\partial n_{2,\bk}}{\partial \xi_{2,\bk}}\mathbf{v}_{2,\bk}\cdot\boldsymbol{\nabla}T\,\frac{\xi_{2,\bk}}{T} &=& -\mathcal{I}^{22}[g_2]-\mathcal{I}^{21}[g_1,g_2], \nn\\\label{eq:BE_thermal_2}
\end{eqnarray}
\end{subequations}
where $\mathcal{I}^{12}$ and $\mathcal{I}^{21}$ are given by Eq.~(\ref{eq:collb}) and the intraband collision integrals are given by
\begin{widetext}
	\begin{eqnarray}
	\mathcal{I}^{ii}[g_j] = \int_{\bk}\int_{\bp'}\int_{\bk'}&&W^{jj}_{\bp\bk\to\bp'\bk}n_{j,\bp}n_{j\bk}(1-n_{j\bp'})(1-n_{j\bk'})
	\left[g_{j\bp}+g_{j{\bk}}-g_{j\bp'}-g_{j\bk'}\right]
	\delta(\varepsilon_{j\bp}+\varepsilon_{j\bk}-\varepsilon_{j\bp'}-\varepsilon_{j\bk'}) \nonumber \\
	&&\times\delta(\bp+\bk-\bp'-\bk'),\;j=1,2;
\end{eqnarray}
\end{widetext}
and $W^{jj}_{\bp\bk\to\bp'\bk'}$ is the probability of intraband scattering. In the forward-scattering limit, the intraband probability is equal to the interband one (and also depends only on $\bq$). However, we will keep $W^{11}_{\bp\bk\to\bp'\bk'}=W^{11}(q)$ and $W^{22}_{\bp\bk\to\bp'\bk'}=W^{22}(q)$ to be different from each other and also from the interband scattering probability $W(q)$  for the sake of generality.   
\begin{comment}
Then the collision integrals can be rewritten as
\begin{widetext}
\begin{subequations}
	\begin{eqnarray}
	&&\mathcal{I}^{ii}[g_j] = \int_{\bk}\int_{\mathbf{q}}\int d\omega W_{ji} (\mathbf{q})
	%\nn\\
	%&&
	\times n_{j\bp}n_{j\bk}(1-n_{j\bp-\bq})(1-n_{j\bk+\bq})
	%\nonumber \\
	%&&\times
	\left[g_{j\bp}+g_{j\bk}-g_{j\bp-\bq}-g_{j\bk+\bq}\right]
	% \nonumber \\
	%&&\times
	\delta(\varepsilon_{j\bp}-\varepsilon_{j\bp-\bq})\delta(\varepsilon_{j\bk}-\varepsilon_{j\bk+\bq}),\nn\\
	\label{colla}\\
	&&\mathcal{I}^{12}[g_1,g_2] = \int_{\bk}\int_{\mathbf{q}}\int d\omega W_{12}(\mathbf{q})n_{1,\bp}n_{2,\bk}(1-n_{1,\bp-\bq})(1-n_{2,\bk+\bq})
	%\nonumber \\&&\times 
	\left[g_{1,\bp}+g_{2,\bk}-g_{1,\bp-\bq}-g_{2,\bk+\bq})\right] 
	%\nonumber \\&&\times
	\delta(\varepsilon_{1,\bp}-\varepsilon_{1,\bp-\bq})\delta(\varepsilon_{2,\bk}-\varepsilon_{2,\bk+\bq}),\nn\\
	\label{collb}
\end{eqnarray}
\end{subequations}
\end{widetext}
where it is understood that $n_{j,{\bf l}\pm \bq}=n_j(\vare_{j,{\bf l}}\pm\omega)$.
\end{comment}
%the energy transfer in the delta functions are dropped as the system is in the FL regime, and it is even valid near a quantum phase transition(QPT).

The solutions for $g_{j,{\bf l}}$ may be sought in the following form 
\begin{eqnarray}
g_{j,\bl}=-\frac{1}{T}(\mathbf{v}_{j,\bl}\cdot\boldsymbol{\nabla}T)\psi_{j}\left(\frac{\xi_{j,\bl}}{T}\right), \label{eq:psi_kappa}
\end{eqnarray}
where $\psi_{j}$ is an odd function of its argument.\footnote{
% to be solved. Rigorously 
Strictly speaking, the even part of $\psi_{j}$ is also to be found as it ensures the boundary condition of no macroscopic current. However, the even part of $\psi_{j}$ leads to a contribution to thermal conductivity which is smaller by a factor of $(T/\varepsilon_F)^2$ compared to that from the odd part~\cite{abrikosov:book} }

In the forward-scattering approximation, $\psi_{j}(x)$ is assumed to be a slowly varying function of its argument. Since $\psi_j(x)$ is odd, the minimal {\em Ansatz} consistent with this requirement is a linear form
\begin{eqnarray}
	\psi_{j}(x)=b_{j} x, \label{lin}
\end{eqnarray} 
where $b_{j}$ are the constants to be determined. Substituting the corresponding expressions $g_{j,\bl}$'s into Eqs.~(\ref{eq:BE_thermal_1}) and (\ref{eq:BE_thermal_2}), multiplying Eq.~(\ref{eq:BE_thermal_1}) [Eq.~(\ref{eq:BE_thermal_2})] by $\xi_{1\bp} \mathbf{v}_{1,\bp}\cdot\boldsymbol{\nabla}T$ [$\xi_{2,\bk}\mathbf{v}_{2,\bk}\cdot\boldsymbol{\nabla}T$], and integrating over $\bp$ ($\bk$), we arrive at a  $2\times 2$ system for $b_{1,2}$, whose solution is  
\begin{subequations}
	\begin{eqnarray}
		b_1 &=& \frac{20\pi^2/m_1T^2}{m^2_1\int dqW^{11}(q)+m^2_2\int dqW(q)}, \label{eq:c_kappa_1}\\
		b_2 &=& \frac{20\pi^2/m_2T^2}{m^2_2\int dqW^{22}(q)+m^2_1\int dqW(q)}.\label{eq:c_kappa_2}
	\end{eqnarray}
\end{subequations}

Once  the non-equilibrium distribution functions are obtained, the thermal current can be  found as 
\begin{eqnarray}
	\mathbf{j}_q &=& 2\sum_{j=1,2} \int\frac{d^3 p_j}{(2\pi)^3}\mathbf{v}_{j,\bp}\xi_j(-Tn^{\prime}_j g_{j,\bp}), \nonumber\\
	&=& \frac{\pi^2nT}{3}\left(\frac{b_1}{m_1}+\frac{b_2}{m_2}\right)\boldsymbol{\nabla}T. \label{eq:energycurrent}
\end{eqnarray}
Substituting Eqs.~(\ref{eq:c_kappa_1}) and (\ref{eq:c_kappa_2}) into Eq.~(\ref{eq:energycurrent}), we obtain the thermal conductivity
\begin{eqnarray}
	\kappa &=& C\frac{np_F}{T}\Big[\frac{1}{m^4_1\int dq W^{11}(q)+m^2_1m^2_2\int dqW(q)} \nonumber \\
	&&+\frac{1}{m^4_2\int dq W^{22}(q)+m^2_1m^2_2\int dqW(q)}\Big]\label{eq:kappa_forward}
\end{eqnarray}
with
$C=20\pi^4/3$.

However, one can check that the higher-order terms in the Taylor series for $\psi_j(x)$ modify the result in Eq.~(\ref{eq:kappa_forward}) in a non-perturbative manner. For example, taking into account the cubic term in the series modifies the numerical prefactor to
$C=(20\pi^4/3)\left[1+7/17\pi^2\right]$.
Higher-order terms will bring additional corrections. Therefore, in contrast to the case of the electrical conductivity considered in Sec.~\ref{sec:rho_for}, the forward-scattering approximation does not produce an asymptotically exact result for the thermal conductivity. However, the higher-order corrections happen to be numerically small: For example, the cubic term changes the prefactor only by $4\%$.
We thus see that approximating $\psi(x)$ by a linear function is still a satisfactory albeit not controllable approximation. In Sec.~\ref{sec:3} we will see why the exact solutions for the electrical and thermal conductivities differ in the forward-scattering limit.

For the screened Coulomb interaction with $\varkappa\ll p_F$, $W^{11}(q)=W^{22}(q)=W(q)$ with $W(q)$ given by Eqs.~(\ref{W_q}) and (\ref{Vq}), and Eq.~(\ref{eq:kappa_forward}) is reduced to
\begin{eqnarray}
	\kappa=C_1\frac{np_F\varkappa^3\epsilon_0^2}{
	Tm^2_1m^2_2e^4}, \label{eq:kappa_forward_1}
\end{eqnarray}
with $C_1=5/12$.
 
In 2D, the corresponding average of the scattering probability
\begin{equation}
	\int \frac{dq}{q} \frac{1}{1-(q/2p_F)^2}W(q),
\end{equation} 
which follows from a derivation similar to Eq.~(\ref{eq:2pF}), is logarithmically divergent both at $q=0$ and $q=2p_F$. However, the corresponding logarithmic factors come with $W(0)$ and $W(2p_F)$, respectively. Since $W(0)\gg W(2p_F)$ for the case of forward scattering, the $q=2p_F$ singularity is subleading compared to the $q=0$ singularity even after resumming the series for the Cooper channel. Therefore, the $2p_F$ singularity can be ignored, and the asymptotic form of $\kappa$ coincides with that found in Ref.~\onlinecite{lyakhov:2003}:
\bea
\kappa\propto \frac{1}{T\ln T}.
\eea

\section{Exact results for the transport coefficients}\label{sec:3}
\subsection{General case}
In this section we obtain exact results for the thermal and electrical conductivities by solving a system of coupled BEs for an arbitrary interaction potential.  (The word ``exact'' here means that the solutions are valid for an arbitrary scattering probability but still only in the limit $T\ll \vare_F$.) We follow the method of solving the integral BE for a single-band FL developed by Abrikosov and Khalatnikov,\cite{abrikosov:1959} Sykes and Brooker,\cite{Sykes,Sykes_2} and Smith, Jensen and Wilkins.\cite{Wilkins} The same formalism was employed by Maldague and Kukkonen,\cite{Kukkonen} who found the electrical resistivity of a CM by a variational solution of the coupled BEs. To the best of our knowledge, however, the thermal conductivity has not been calculated, and thus the Lorentz ratio has not been determined. In what follows, we obtain exact results both for the electrical and thermal conductivities, and thus for the Lorentz ratio.    
\begin{figure}[h]
        \includegraphics[height=1.5in,width=3in]{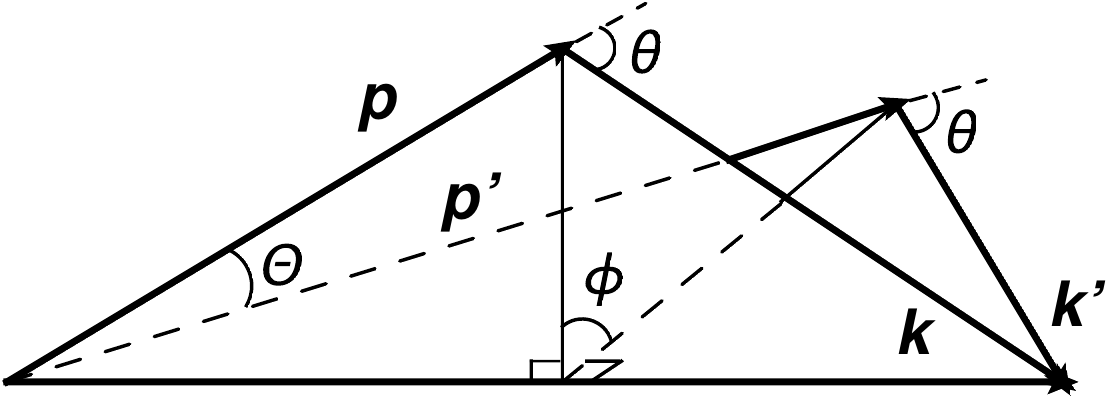}
		\caption{Schematics of a collision process with incoming momenta $\bp, \bk$ and outgoing momenta $\bp^\prime, \bk^\prime$.} 
		\label{fig:fig1}
\end{figure}
Following Refs.~\onlinecite{abrikosov:1959,Sykes,Sykes_2,Wilkins},  we rewrite the intra- and interband collision integrals as
\begin{widetext}
\begin{subequations}
	\begin{eqnarray}
	\mathcal{I}^{jj}[g_j] = \frac{m^3_j}{8\pi^4}\int d\varepsilon_\bk\int &&d\omega n_F(\varepsilon_\bp)n_F(\varepsilon_\bk)\left[1-n_F(\varepsilon_{\bp}-\omega)\right]\left[1-n_F(\varepsilon_{\bk}+\omega)\right] \nonumber \\
		&&\times \int \frac{d\Omega}{4\pi}\int^{2\pi}_0 \frac{d\phi_\bk}{2\pi}\frac{W^{jj}(\theta,\phi)}{\cos(\theta/2)}\left(g_{j,\bp}+g_{j,\bk}-g_{j,\bp'}-g_{j,\bk'}\right),\,j=1,2; \\ \nonumber\\
	\mathcal{I}^{12}[g_1,g_2] = \frac{m_1 m^2_2}{8\pi^4}\int d\varepsilon_\bk\int &&d\omega n_F(\varepsilon_\bp)n_F(\varepsilon_\bk)(1-n_F(\varepsilon_\bp-\omega))(1-n_F(\varepsilon_\bk+\omega)) \nonumber \\
		&&\times \int \frac{d\Omega}{4\pi}\int^{2\pi}_0 \frac{d\phi_\bk}{2\pi}\frac{W(\theta,\phi)}{\cos(\theta/2)}\left[g_{1, \bp}+g_{2, \bk}-g_{1, \mathbf{p}'}-g_{2, \mathbf{k}'}\right],
\end{eqnarray}
\end{subequations}
\end{widetext}
where the angles are defined as follows (see Fig.~\ref{fig:fig1}): $\theta$ is the angle between the initial state momenta $\bp$ and $\bk$, $\phi$ is the angle between the planes formed by  $\bp$ and $\bk$, and the final state momenta, $\bp'$ and $\bk'$, respectively, $\phi_\bk$ is the azimuthal angle of $\bk$ relative to $\bp$, and $d\Omega=d\theta \sin\theta d\phi$. As before, $W^{jj}(\theta,\phi)$ with $j=1,2$ and $W(\theta,\phi)$ are the intra- and interband scattering probabilities, respectively, while ${\cal I}^{21}[g_1,g_2]$ is obtained from ${\cal I}^{21}[g_1,g_2]$ by interchanging indices $1$ and $2$.

We again seek $g_{j,\bl}$ in the forms of Eqs.~(\ref{eq:psi_rho}) and (\ref{eq:psi_kappa}) for the electrical and thermal conductivities, respectively.   This leads to a system of integral equations for the unknown energy-dependent functions $\varphi_{j}(x)$ and $\psi_{j}(x)$, which can be solved by converting the integral equations into  a system of second-order differential equations for Fourier transforms of the distribution functions. For brevity, we provide here only the final results delegating the computational details to Appendix \ref{sec:app}. First, we introduce some definitions: 
\begin{subequations}
\begin{eqnarray}
\tau_{\rho 1}^{-1} &=& \frac{m_1 m_2^2 T^2}{8\pi^4}\Big\langle \frac{W(\theta,\phi)}{\cos(\theta/2)}\Big\rangle,\label{tk1}\\
\lambda_\rho &=& \Big\langle \frac{W(\theta,\phi)\cos\Theta}{\cos(\theta/2)}\Big\rangle\Big/\Big\langle \frac{W(\theta,\phi)}{\cos(\theta/2)}\Big\rangle,\label{lr}\\
    \tau^{-1}_{\kappa 1} &=& \frac{m_1T^2}{8\pi^4}\left[m^2_1\Big\langle \frac{W^{11}(\theta,\phi)}{\cos(\theta/2)} \Big\rangle+m^2_2\Big\langle \frac{W(\theta,\phi)}{\cos(\theta/2)}\Big\rangle\right], \nn\\\label{tk2}\\
    \lambda_{\kappa 1} &=& \frac{m^2_1\Big\langle \frac{W^{11}(\theta,\phi)(1+2\cos\theta)}{\cos(\theta/2)}\Big\rangle+m^2_2\Big\langle \frac{W(\theta,\phi)\cos\Theta}{\cos(\theta/2)}\Big\rangle}{m^2_1\Big\langle \frac{W^{11}(\theta,\phi)}{\cos(\theta/2)} \Big\rangle+m^2_2\Big\langle \frac{W(\theta,\phi)}{\cos(\theta/2)}\Big\rangle}, \nn\\
    \label{lk1}\\
	\beta_{\kappa 1} &=& \frac{m_1m_2\Big\langle \frac{W(\theta,\phi)(1+2\cos\theta-\cos\Theta)}{\cos(\theta/2)}\Big\rangle}{m^2_1\Big\langle \frac{W^{11}(\theta,\phi)}{\cos(\theta/2)} \Big\rangle+m^2_2\Big\langle \frac{W(\theta,\phi)}{\cos(\theta/2)}\Big\rangle}, \label{bk1}
	\end{eqnarray}
\end{subequations}
where $\Theta$ is the scattering angle related to the angles $\theta$ and $\phi$ via 
$
\sin(\Theta/2) = \sin(\theta/2)\sin(\phi/2)
$
 and $\langle\cdots\rangle$ denotes $\int \frac{d\Omega}{4\pi}\cdots$.
The quantities with index $2$, e.g., $\tau_{\rho 2}$, $\tau_{\kappa 2}$, etc., are obtained from $\tau_{\rho 1}$, $\tau_{\kappa 1}$, etc. by interchanging indices $1$ and $2$ in the right-hand sides of Eqs.~(\ref{tk1}) to (\ref{bk1}). Note that $\tau_{\rho j}$ and $\tau_{\kappa j}$ can be interpreted as the relaxation times of the electrical and thermal currents in the $j\text{th}$ band, respectively, while $\lambda_\rho$ is an average cosine of the scattering angle.

Let the energy-dependent part of the distribution function in the presence of an electric field be
\bea
\varphi_j(x)=\tau_{\rho j} \cosh\left(\frac{x}{2}\right) \Phi_j(x),\; j=1,2.\label{phiPhi}
\eea
 To find the electrical current, one needs to know only the sum $\Phi(x)\equiv\Phi_1(x)+\Phi_2(x)$, whose Fourier transform 
$\tilde\Phi(k)=\int dx e^{ikx} \Phi(x)$ is given by
\begin{eqnarray}
\tilde{\Phi}(k)=-\frac{2}{\pi}\sum_{l=0}^\infty\frac{4l+3}{(l+1)(2l+1)}\,\frac{P^1_{2l+1}(\tanh \pi k)}{(l+1)(2l+1)-2\lambda_\rho+1}, 
\nn\\\label{phik}
\end{eqnarray}
where $P_{l}^{m}(x)$ are the associated Legendre polynomials.

Using Eqs.~(\ref{eq:psi_rho}), (\ref{phiPhi}) and (\ref{phik}), we find the electrical resistivity as 
\begin{eqnarray}
	\rho = \frac{m^2_1m^2_2T^2}{8\pi^2e^2n} \Big\langle \frac{W(\theta,\phi)}{\cos(\theta/2)}\Big\rangle {\cal P}\left(2(1-\lambda_\rho)\right), \label{eq:rho_2}
\end{eqnarray}
where
	\bea
 \frac{1}{{\cal P}(x)}&=&
	%\Big[
	\sum^\infty_{l=0} \frac{4l+3}{(l+1)(2l+1)}
	% \nonumber \\
	%&&\times
	\frac{1}{(l+1)(2l+1)-1+
	%(2\lambda_\rho-1)
	x}\nn\\
	%\Big]^{-1}. \label{eq:rho_2}
	&=&\frac{1}{2(x-1)}\Big[\gamma+\ln 2 + \frac{1}{2}D_\Gamma\left(\frac{3}{4}+\frac{1}{4}\sqrt{9-8x}\right) \nonumber\\
	&+&\frac{1}{2}D_\Gamma\left(\frac{3}{4}-\frac{1}{4}\sqrt{9-8x}\right)\Big],
	\label{eq:curl_P_x}
	\eea
$\gamma\approx 0.58$ is the Euler constant and $D_\Gamma(x)\equiv d\ln\Gamma(x)/dx$ is the digamma function. Note that the zero of the denominator at $x=1$ in ${\cal P}(x)$ is compensated by the vanishing of the numerator. Also,  ${\cal P}(x)$ remains real for $x>9/8$, when the arguments of the square roots become negative.
	
Likewise, let the energy-dependent part of the distribution function
 in the presence of a thermal gradient be
\bea\psi_j(x)=\tau_{\kappa j} \cosh(x/2) \Psi_j(x)\label{psiPsi}.
\eea 
To find the thermal current, we need to know $\Psi_1(x)$ and $\Psi_2(x)$ individually. Their Fourier transforms are given by
\bwt
\begin{subequations}
\bea
\tilde{\Psi}_1(k)&=&-i\sum_{l=0}^\infty \frac
{(l+1)(2l+3)-\lambda_{\kappa 2}-\beta_{\kappa 1} \tau_{\kappa 2}/\tau_{\kappa 1}}
{\left[(l+1)(2l+3)-\lambda_{\kappa 1}\right]
\left[(l+1)(2l+3)-\lambda_{\kappa 2}\right] -\beta_{\kappa 1}\beta_{\kappa 2}
}\frac{4l+5}{(l+1)(2l+3)} P_{2l+2}^1(\tanh \pi k), \label{psi1}\\
\tilde{\Psi}_2(k)&=&-i\sum_{l=0}^\infty \frac
{(l+1)(2l+3)-\lambda_{\kappa 1}-\beta_{\kappa 2} \tau_{\kappa 1}/\tau_{\kappa 2}}
{\left[(l+1)(2l+3)-\lambda_{\kappa 1}\right]
\left[(l+1)(2l+3)-\lambda_{\kappa 2}\right] -\beta_{\kappa 1}\beta_{\kappa 2}
}\frac{4l+5}{(l+1)(2l+3)} P_{2l+2}^1(\tanh \pi k).\label{psi2}
\eea
\end{subequations}
\ewt

The thermal conductivity is also given by an infinite series: 
\begin{eqnarray}
	\kappa &=& \frac{4\pi^4n/m^2_1T}{m^2_1\Big\langle W^{11}(\theta,\phi)/\cos\frac{\theta}{2}\Big\rangle+m^2_2\Big\langle W(\theta,\phi)/\cos\frac{\theta}{2}\Big\rangle} \nonumber \\
	&&\times \sum^\infty_{l=0} \frac{4l+5}{(l+1)(2l+3)}\times \nonumber \\
	&& \frac{(l+1)(2l+3)-\lambda_{\kappa 2}-\beta_{\kappa 1}\tau_{\kappa 2}/\tau_{\kappa 1}}{\left[(l+1)(2l+3)-\lambda_{\kappa 1}\right]\left[(l+1)(2l+3)-\lambda_{\kappa 2}\right]-\beta_{\kappa 1}\beta_{\kappa 2}} \nonumber \\
	&& +(m_1\to m_2). \label{eq:kappa_2} 
\end{eqnarray}   
In contrast to the electrical resistivity, the series for $\kappa$ cannot be reduced to a more compact form in terms of the digamma function.

If the interband scattering probability, $W(\theta,\phi)$, is set to zero and the band masses are taken to be the same, Eq.~(\ref{eq:kappa_2}) is reduced to (twice) the thermal conductivity of a single-band FL~\cite{Sykes,Wilkins}. In this case, $\beta_{\kappa 1}$ and $\beta_{\kappa 2}$ vanish, while $\lambda_{\kappa 1}$ and $\lambda_{\kappa 2}$ become identical to $\lambda_K$ of Ref.~\onlinecite{Sykes} and to $\alpha/2$ of Ref.~\onlinecite{Wilkins}.

\subsection{Limiting cases}
Eqs.~(\ref{eq:rho_2}) and (\ref{eq:kappa_2}) are valid for arbitrary forms of the scattering probabilities $W^{11}(\theta,\phi)$, $W^{22}(\theta,\phi)$, and $W(\theta,\phi)$. It is instructive, however, to analyze the limiting cases of forward and isotropic scattering, which are considered in this section. 
\subsubsection{Forward scattering}

In the forward-scattering limit,  the parameter $\lambda_\rho$ in Eq.~(\ref{lr}) can be written as
\begin{eqnarray}
	\lambda_\rho =1-\frac{\overline{\Theta^2}}{2},
	\eea
	where
	\bea
\overline{\Theta^2}\equiv\Big\langle \frac{W(\theta,\phi)\Theta^2}{\cos(\theta/2)}\Big\rangle\Big/\Big\langle \frac{W(\theta,\phi)}{\cos(\theta/2)}\Big\rangle \ll 1
\end{eqnarray}
is the average square of the scattering angle. The $l=0$ term in the series for $\tilde \Phi(k)$ in Eq.~(\ref{phik}) is singular in the limit of $\lambda_\rho\to 1$, while the rest of the terms are regular.  Keeping only the $l=0$ term, we find 
\bea
\tilde\Phi(k)= -\frac{6}{\pi \overline{\Theta^2}} P^1_1(\tanh \pi k)=\frac{6}{\pi \overline{\Theta^2}}\frac{1}{\cosh \pi k}
\eea
and thus
\bea
\Phi(x)=\int^\infty_{-\infty}\frac{dk}{2\pi} \tilde\Phi(k)= \frac{3}{\pi^2\overline{\Theta^2}}\frac{1}{\cosh\frac x2}.
\eea
Recalling 
relation (\ref{phiPhi}), we see that $\varphi(x)=\varphi_1(x)+\varphi_2(x)$ is independent of $x$:
\bea
\varphi(x)=\left(\tau_{\rho 1}+\tau_{\rho 2}\right)\frac{3}{\pi^2\overline{\Theta^2}}.\label{const}
\eea
This means that the non-equilibrium part of the distribution function in Eq.~(\ref{eq:psi_rho}) is indeed almost independent of energy, in agreement with the argument given in Sec.~\ref{sec:rho_for}. The energy dependence of the distribution function results from  the remainder of the series in Eq.~(\ref{phik}) with $l\geq 1$. In all terms in this remainder one can safely set $\lambda_\rho=1$. The result is some function of $k$, which is parametrically smaller than the $l=0$ term by $\overline{\Theta^2}\ll 1$. As result, a correction to 
Eq.~(\ref{const}) is some (even) function of $x$, which is ${\cal O}(1)$ for $x\sim 1$.

Substituting $\lambda_\rho$ into Eq.~(\ref{eq:rho_2}), we find the electrical resistivity as
\begin{eqnarray}
	 \rho=\frac{m^2_1m^2_2T^2}{24\pi^2e^2n} \Big\langle \frac{W(\theta,\phi)}{\cos(\theta/2)} \Theta^2\Big\rangle, \label{eq:rho_3}
\end{eqnarray}
where we have used that ${\cal P}(x)\approx x/3$ for $x \ll 1$. 
For a generic value of the angle $\theta$ between the initial momenta $\bk$ and $\bp$, a small value of $\Theta$ can only be achieved if the angle $\phi$ between the planes formed by the initial and final momenta is small. Then $\Theta\approx \sin(\theta/2)\phi$. If $W$ depends only on $\Theta$ or, which is the same, on the momentum transfer $q=2p_F\sin\Theta/2\approx p_F\Theta$, the average  in Eq.~(\ref{eq:rho_3}) can be written as
\bea
\Big\langle \frac{W(\theta,\phi)\Theta^2}{\cos(\theta/2)}\Big\rangle=\frac{1}{2p_F^3} \int dq q^2 W(q).
\eea
Substituting the last result into Eq.~(\ref{eq:rho_3}), we see that it  indeed coincides with Eq.~(\ref{eq:rho_forward}) for $\rho$ obtained in Sec.~\ref{sec:rho_for}.

We now turn to the thermal conductivity. In the forward-scattering limit, $W^{11}(\theta,\phi)$, $W^{22}(\theta,\phi)$ and $W(\theta,\phi)$ are equal because the exchange term can be neglected. 
Then the parameters entering Eqs.~(\ref{psi1}) and (\ref{psi2}) can be simplified as 
\begin{eqnarray}
    && \lambda_{\kappa 1} \approx 1+ \frac{2m^2_1}{m^2_1+m^2_2}\overline{\cos\theta}, \; \lambda_{\kappa 2} \approx 1+ \frac{2m^2_2}{m^2_1+m^2_2}\overline{\cos\theta},\nn \\
    &&\beta_{\kappa 1}=\beta_{\kappa 2} \approx \frac{2m_1m_2} {m^2_1+m^2_2}\overline{\cos\theta}, \;  \frac{\tau_{\kappa 2}}{\tau_{\kappa 1}}\approx \frac{m_1}{m_2},\label{paramkappa}
\end{eqnarray}
where
\begin{eqnarray}
\overline{\cos\theta} & \equiv &\Big\langle \frac{W(\theta,\phi)\cos\theta}{\cos(\theta/2)}\Big\rangle \Big/ \Big\langle \frac{W(\theta,\phi)}{\cos(\theta/2)}\Big\rangle.
\end{eqnarray}
However, because generic values of these parameters are of order one, the series in Eqs.~(\ref{psi1}) and (\ref{psi2}) cannot be simplified any further.  This implies that the energy-dependent part of the distribution functions in the presence of a thermal gradient, $\psi_{j}(x)$, are some odd functions of $x$ which, generally speaking, cannot be approximated by a linear form of Eq.~(\ref{lin}).

However, substituting Eq.~(\ref{paramkappa}) into the series for the thermal conductivity in Eq.~(\ref{eq:kappa_2}), we find  that all physical parameters drop out, and the series is reduced to a simple number: 
\bea
\sum^\infty_{l=0} \frac{4l+5}{(l+1)(2l+3)}\frac{1}{(l+1)(2l+3)-1}=1.
\eea
Hence the thermal conductivity in the forward-scattering limit is given by
\begin{eqnarray}
	\kappa=\frac{4\pi^4n}{m^2_1m^2_2T}\frac{1}{\Big\langle W(\theta,\phi)/\cos(\theta/2)\Big\rangle}. \label{eq:kappa_3}
\end{eqnarray}
With $\Big\langle W(\theta,\phi)/\cos(\theta/2)\Big\rangle=(1/2p_F)\int dq W(q)$, the last result can be rewritten as
\begin{eqnarray}
	\kappa=
	\frac{8\pi^4n p_F}{m^2_1m^2_2T}\frac{1}{\int dq W(q)}. \label{eq:kappa_4}
\end{eqnarray}
For $W^{11}(q)=W^{22}(q)=W(q)$, the approximate result for $\kappa$ in Eq.~(\ref{eq:kappa_forward}) is reduced to the same form as that in 
Eq.~(\ref{eq:kappa_4}) but with a numerical prefactor $C=20\pi^4/3$, which differs from the exact prefactor of $8\pi ^4$ in Eq.~(\ref{eq:kappa_4})  by $17\%$. Keeping the cubic term in $\psi_{1,2}(x)$ reduces the disagreement to $13\%$, etc. Therefore, an approximate method of  Sec.~\ref{sec:kappa_for} gives a reasonably accurate albeit not asymptotically exact result for $\kappa$.

%In order to estimate the numeric value of the Lorentz number we take the screened Coulomb potential in Eq.~(\ref{eq:W_q}) with $q \approx p_F\sin(\theta/2)\phi$ for $W^{11}(\theta,\phi)$, $W^{22}(\theta,\phi)$ and $W(\theta,\phi)$. Evaluating the angular average of the scattering probability we find
%\bea
%\Big\langle \frac{W(\theta,\phi)\Theta^2}{\cos(\theta/2)}\Big\rangle\approx \frac{8\pi^4e^4}{\epsilon^2_\infty p^3_F\varkappa}, \\
%\Big\langle \frac{W(\theta,\phi)}{\cos(\theta/2)}\Big\rangle\approx \frac{8\pi^4e^4}{\epsilon^2_\infty p_F\varkappa^3}.
%\eea 
In the case of the screened Coulomb interaction, the result for $\rho$ remains the same as that in Eq.~(\ref{eq:rho_forward_1}), while 
$\kappa$ is given by
\begin{eqnarray}
	%\rho & \approx & \frac{\pi^2}{3e^2}\frac{T^2 m^2_1m^2_2e^4}{n
	%6
	%\epsilon_\infty^2 p^3_F \varkappa}, \label{eq:rho_forward_2}\\
	\kappa= \frac{np_F\varkappa^3\epsilon_0^2}{
	%6
	2Tm^2_1m^2_2e^4}. \label{eq:kappa_forward_2}
\end{eqnarray}
\subsubsection{Isotropic scattering}
Another limiting case is isotropic scattering, which corresponds to a short-range (Hubbard-like) interaction.  In this case, the intra- and interband scattering probabilities reduce to three constants: $W^{11}$, $W^{22}$, and $W$, respectively. The angular averages of the scattering probability can then be readily performed, and we find $\beta_{\kappa 1}=\beta_{\kappa 2}=0, \lambda_{\kappa 1}=\lambda_{\kappa 2}=\lambda_{\rho}=1/3$. Substituting the above parameters into Eqs.~(\ref{eq:rho_2}) and (\ref{eq:kappa_2}),  we find
\begin{subequations}
\bea
    \rho&=& 0.0050\frac{m^2_1m^2_2T^2}{e^2n} W, \label{eq:rho_isotropic}\\
	\kappa &= & 303.9\frac{n}{Tm^2_1m^2_2W}\left(\frac{1}{\frac{m^2_1}{m^2_2}\frac{W^{11}}{W}+1}+\frac{1}{\frac{m^2_2}{m^2_1}\frac{W^{22}}{W}+1}\right).\nn\\ \label{eq:kappa_isotropic}
\eea
(The odd-looking numerical prefactors in these equations result from numerical summation of the corresponding series.)
If $W^{11}=W^{22}=W$, the result for $\kappa$ is reduced to $\kappa=  303.9n/Tm^2_1m^2_2W$. If $W\ll W^{11},\,W^{22}$, $\kappa$ is reduced to the sum of two single-band conductivities: $\kappa\approx (303.9n/T)\sum_{j=1,2}1/(m^4_jW^{jj})$. 
\end{subequations}
\section{Lorentz ratio}
\label{sec:lorentz}
\subsubsection{Forward scattering}
For clarity, we again assume that $W^{11}(\theta,\phi)=W^{22}(\theta,\phi)=W(\theta,\phi)$. Using Eqs.~(\ref{eq:rho_3}) and (\ref{eq:kappa_3}), we find a simple result for the Lorentz ratio
\begin{eqnarray}
	\frac{L}{L_0}= \frac{\overline{\Theta^2}}{2},\label{L_for}
\end{eqnarray}
where $L_0$ is the Sommerfeld constant given by Eq.~(\ref{L0}). Note that this result does not depend on the mass ratio.
By assumption of forward scattering, $\overline{\Theta^2}\ll 1$ and hence $L/L_0$,  in principle, can be arbitrarily small. For the screened Coulomb potential in Eq.~(\ref{Vq}), Eq.~(\ref{L_for}) is reduced to
\begin{eqnarray}
	\frac{L}{L_0}= \frac{\varkappa^2}{2p^2_F}.\label{LC}
\end{eqnarray}

\subsubsection{Isotropic scattering}
Using Eqs.~(\ref{eq:rho_isotropic}) and (\ref{eq:kappa_isotropic}), we find the Lorentz ratio for the isotropic-scattering case
\begin{eqnarray}
	\frac{L}{L_0}=0.46 \left(\frac{1}{\frac{m^2_1}{m^2_2}\frac{W_{11}}{W}+1}+\frac{1}{\frac{m^2_2}{m^2_1}\frac{W_{22}}{W}+1}\right).
	\label{iso}	
\end{eqnarray}
If $W^{11}=W^{22}=W$, the Lorentz ratio  assumes a universal value of $L/L_0= 0.46$, which does not depend on the mass ratio. If $W\ll W^{11},W^{22}$, the Lorentz ratio is reduced in proportion to the ratios of the inter- and intraband scattering probabilities:
\bea
\frac{L}{L_0} =0.46 \left(\frac{W}{W^{11}}\frac{m^2_2}{m^2_1}+\frac{W}{W^{22}}\frac{m^2_1}{m^2_2}\right)\ll 1.
\eea
\subsubsection{Exact result}
One can also calculate the Lorentz ratio without assuming either forward or isotropic scattering for a given interaction potential. To be specific, we use again the screened Coulomb potential in Eq.~(\ref{Vq}), but  now assume that $\varkappa/p_F$ can be arbitrary. Although Eq.~(\ref{Vq}) is not, strictly speaking, valid for $\varkappa\gtrsim p_F$, one can still view this equation as a model with an adjustable parameter ($\varkappa/p_F$), which allows one to interpolate between the limits of forward and isotropic scattering. 
For a generic interaction potential, one needs to restore the exchange term in the scattering amplitude, as given by Eq.~(\ref{sca}). To minimize the number of free parameters, however, we assume that the distance between the centers of the electron and hole FSs [$p_0$ in Eq.~(\ref{bands})] is much larger than $\varkappa$, so that the exchange term in the interband scattering amplitude can still be neglected, but take into account the exchange term in the intraband amplitude.
We then calculate the parameters in Eqs.~(\ref{tk1}-\ref{bk1}) and use the exact results for $\rho$ and $\kappa$, Eqs.~(\ref{eq:rho_2}) and (\ref{eq:kappa_2}), respectively. 

\begin{figure}[h]
        \includegraphics[height=1.5in,width=3in]{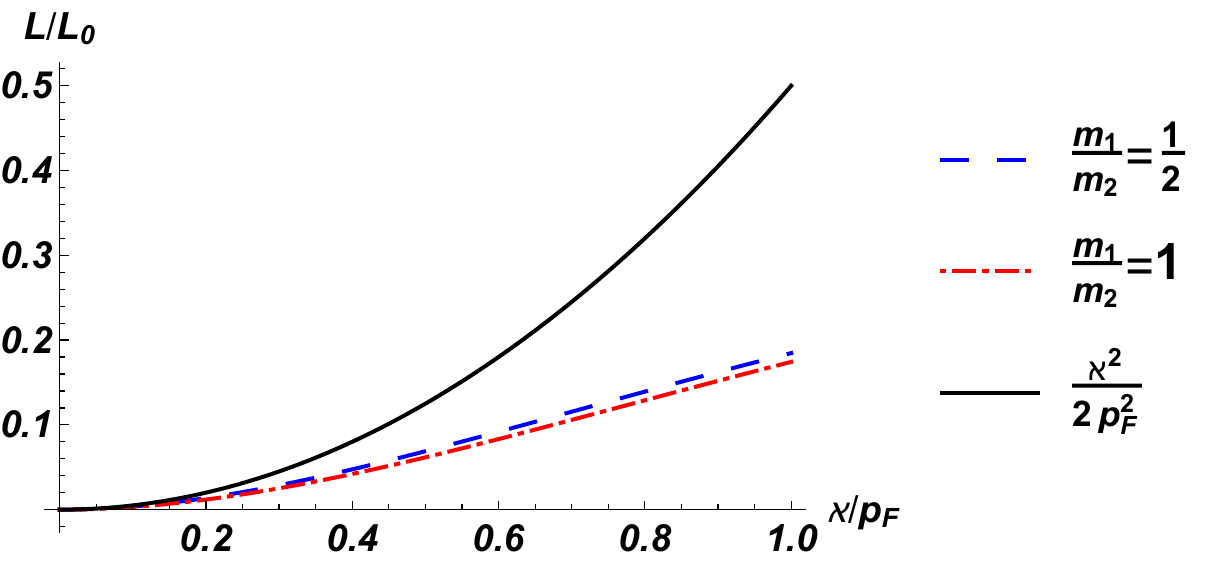}
         \includegraphics[height=1.5in,width=3in]{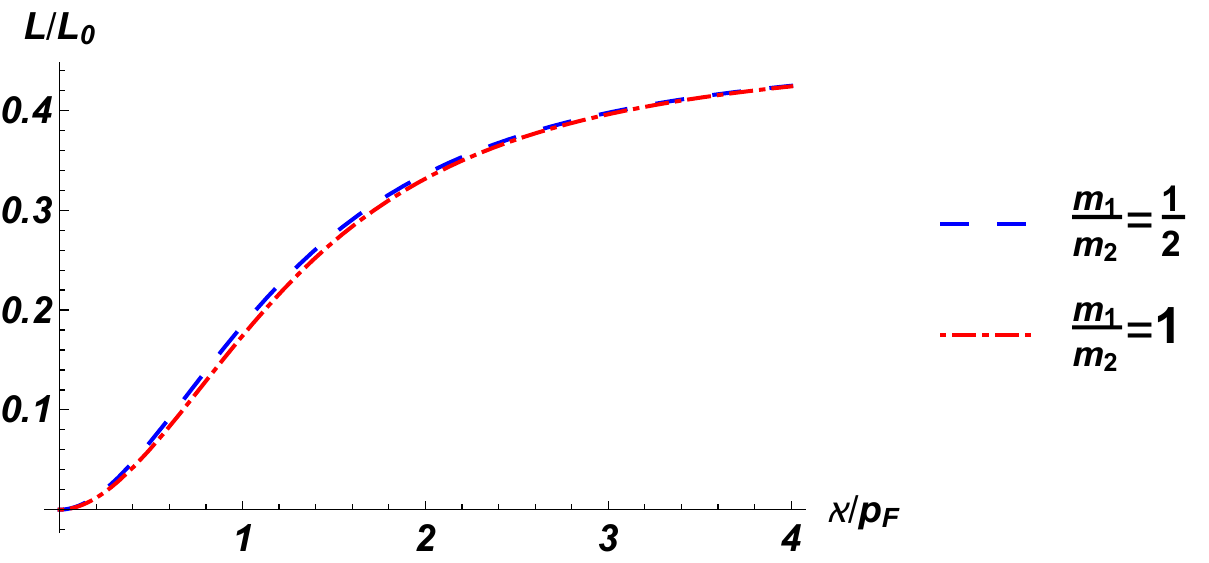}
		\caption{\label{fig:Lofkappa} The Lorentz ratio in units of the Sommerfeld constant as a function of $\varkappa/p_F$, where $\varkappa$ is the inverse screening length defined by Eq.~(\ref{Vq}).} 
		\label{fig:fig2}
\end{figure}

The result of this calculation is shown in the top  panel of Fig.~\ref{fig:Lofkappa} for 
$0\leq \varkappa/p_F\leq 1$ and in the bottom panel for  $0\leq \varkappa/p_F\leq 4$.  The dashed and dashed-and-dotted lines in both panels correspond to exact solutions for $m_1/m_2=1/2$ and $m_1/m_2=1$, respectively. The solid line in the top panel depicts the forward-scattering limit given by Eq.~(\ref{LC}). We see that the exact result matches the approximate one for $\varkappa/p_F\lesssim 0.2$ but goes below the approximate one for larger $\varkappa/p_F$. The dependence of the exact result on the mass ratio, which is absent in the forward-scattering limit, remains very weak for arbitrary $\varkappa/p_F$. For larger values of $\varkappa/p_F$, the Lorentz ratio shows a clear tendency to saturation This agrees with the analytic result in the isotropic-scattering limit [cf. Eq.~(\ref{iso})] because our current model corresponds to $W^{11}=W^{22}=W$. The value at saturation is also quite close to the analytic result of $L/L_0=0.46$. 

\subsection{Lorentz ratio of WP$_2$}
The Lorentz ratio measures the relative rates of relaxation of electrical and thermal currents, with $L=L_0$ indicating that the two rates are same. Recent experiments on a compensated metal WP$_2$\cite{Jaoui:2018} found a very low Lorentz ratio ($L/L_0\approx 0.2$), which  indicates that the electrical current is relaxed much slower than the thermal one. This low Lorentz ratio was observed in the temperature range where both $\rho$ and $T/\kappa$ scale as $T^2$, which suggests that the dominant scattering mechanism is electron-electron interaction. According to the discussion in the previous two sections, a low Lorentz ratio can occur if either the interaction is of a long-range type, or  if the interaction is of a short-range type but intraband scattering is much stronger than interband one.

In the first scenario, the Lorentz ratio is parameterized by the average square of the scattering angle [See Eq.~(\ref{L_for})] which, for the screened Coulomb potential translates into Eq.~(\ref{LC}). In a multiband system 
\bea
\varkappa^2 =\frac{4 e^2}{\pi \epsilon_0}\sum_{j} m_j p_{F,j},
\eea
where $m_j$ and $p_{F,j}$ are the effective mass and Fermi momentum of the $j\text{th}$ band, respectively. WP$_2$ has two electron pockets with masses $m_\gamma=0.87\,m_0$ and  $m_\delta=0.99\,m_0$, and two hole pockets with masses $m_\alpha= 1.67\, m_0$ and $m_\beta =0.89\, m_0$ (Ref.~\onlinecite{Gooth:2017}).\footnote{Although our model has only two pockets, additional pockets can be incorporated by replacing $m_1$ and $m_2$ by properly averaged masses. Since the masses cancel out in $L/L_0$, a particular way of averaging does not matter} The total number density of electrons (equal to that of holes) is $n \approx 2.5 \times 10^{21}$cm$^{-3}$.
According to the exact solution of the BEs (dashed and dashed-and-dotted lines in Fig.~\ref{fig:fig2}), the Lorentz ratio reaches the observed value of $0.2$ already at $\varkappa/p_F\approx 1$. With material parameters indicated above, we have
	$\varkappa/p_F\approx \sqrt{30.8/\epsilon_0}$, and the forward-scattering scenario can explain the experiment only if $\epsilon_0$ is quite large: $\epsilon_0\gtrsim 30$. Interestingly, this may be the case for WP$_2$. Indeed, although we are not aware of optical measurements on this material, an anomalously large value of $\epsilon_\infty$ (from $75$ to $91$, depending on orientation) have recently been reported for a cousin material WTe$_2$ (Ref.~\onlinecite{frenzel:2017}), which is also a type-II Weyl semimetal. Since $\epsilon_0\geqslant\epsilon_\infty$, the condition $\epsilon_0\gtrsim 30$ is then satisfied within a good margin. Alternatively, a small Lorentz ratio can result from strong intraband scattering with short-range interaction. With comparable masses of electron and hole bands, the observed value of $L/L_0$ can be already achieved if $W/W^{11}=W/W^{22}\approx 0.4$, which is not unrealistic. Optical data for $\epsilon_\infty$ in WP$_2$ are obviously needed to discriminate between the forward- and isotropic-scattering scenarios.

\section{Conclusions}
\label{concl}
In this paper, we calculated electrical and thermal conductivities of a clean compensated two-band metal with intercarrier interaction as the dominant scattering mechanism. From the theoretical standpoint, it is an attractive toy model which allows one to study both electrical and thermal transport properties, without invoking additional mechanisms of momentum relaxation. However, this model is also relevant to a large class of materials, i.e., metals and semimetals with even number of electrons per unit cell, which include many elemental metals, group V semimetals, graphite, parent states of iron-based superconductors, type-II Weyl semimetals, etc. To find the electrical and thermal conductivities, we solved exactly the system of coupled Boltzmann Equations, describing both inter- and intraband scattering, and analyzed the limiting cases of forward and isotropic scattering. We showed that the forward-scattering limit of the electrical conductivity can be obtained without knowing the exact solution: By assuming from the very beginning that the non-equilibrium part of the distribution function depends primarily on the directions of carriers' momenta but not on their energies. For the thermal conductivity, the same procedure leads to a reasonable albeit not asymptotically exact approximation. We obtained the exact result for the Lorentz ratio and showed that it takes a particularly simple form, parameterized by the average square of the scattering angle,  in the forward-scattering limit [cf.~Eq.~(\ref{L_for})]. We analyzed the Lorentz ratio of a type-II Weyl semimetal WP$_2$ and showed that a strong downward violation of the Wiedemann-Franz Law observed in this material\cite{Jaoui:2018} can be explained within the forward-scattering model, provided that the high-frequency value of the dielectric constant in this material is sufficiently large.

\acknowledgements
We thank I. Aleiner, K. Behnia, I. Mazin, P. Hirschfeld, and A. Subedi for stimulating discussions. S. L. is acknowledges support as a Dirac Post-Doctoral Fellow at the National High Magnetic Field Laboratory, which is supported by the National Science Foundation via Cooperative Agreement No. DMR-1157490, the State of Florida, and the U.S. Department of Energy. D. L. M. acknowledges financial support from the National Science Foundation under Grant NSF-DMR-1720816. 
\appendix
\section{Exact solution of the coupled Boltzmann equations }
\label{sec:app}
\bwt
In the Appendix we provide detailed derivations of the exact results for the electric resistivity and the thermal conductivity. To reiterate, ``exact" here means that the results are valid for an arbitrary scattering probability but only for $T\ll \vare_F$.
\subsection{Electrical resistivity}
To find the electrical resistivity of a two-band compensated metal, we start with the system of coupled Boltzmann equations, Eqs.~(\ref{eq:BE_rho_1}) and (\ref{eq:BE_rho_2}). The non-equilibrium parts of the distribution functions are parameterized by Eq.~(\ref{eq:psi_rho}).
%\begin{eqnarray}
%	-e\mathbf{E}\cdot\mathbf{v}_1\frac{\partial{n}}{\partial\xi_1} &=& -\mathcal{I}^{12}[g_1,g_2], \\
%	-e\mathbf{E}\cdot\mathbf{v}_2\frac{\partial{n}}{\partial\xi_2} &=& -\mathcal{I}^{21}[g_1,g_2].
%\end{eqnarray}
%Since intraband scattering does not affect the electrical current of carriers with parabolic dispersions, the corresponding terms in the collision integrals were dropped. 
%$g_i$ has the form $g_i=-e\mathbf{v}_i\cdot\mathbf{E}\varphi_i(x_i)/T$, where $\varphi_i$ is an even function to be determined. 
Defining dimensionless variables $x_j=\xi_j/T$ and $y=\omega/T$, where $\omega$ is the energy transfer in the scattering processes,
 we rewrite Eqs.~(\ref{eq:BE_rho_1}) and (\ref{eq:BE_rho_2}) in the notations of Ref.~\onlinecite{abrikosov:1959} as
\begin{subequations}
	\begin{eqnarray}
	\frac{1}{m_1}n_F(x_1)\left[1-n_F(x_1)\right] &=& \frac{m_1m^2_2T^2}{8\pi^4}\int \frac{d\Omega}{4\pi} \frac{W(\theta,\phi)}{\cos(\theta/2)} \int dx_2 \, dy \, \mathcal{Q}(x_1,x_2,y)  \nonumber \\
	&& \times\left[\frac{\varphi_1(x_1)}{m_1}-\frac{\varphi_2(x_2)}{m_2}\cos\theta-\frac{\varphi_1(x_1-y)}{m_1}\cos\Theta+\frac{\varphi_2(x_2+y)}{m_2}\cos\theta_{\bk'\bp}\right], \label{eq:BE_5} \\
	\frac{1}{m_2}n_F(x_2)\left[1-n_F(x_2)\right] &=& \frac{m_2m^2_1T^2}{8\pi^4}\int \frac{d\Omega}{4\pi} \frac{W(\theta,\phi)}{\cos(\theta/2)} \int dx_1 \, dy \, \mathcal{Q}(x_1,x_2,y)\nonumber \\
	&&\times \left[\frac{\varphi_2(x_2)}{m_2}-\frac{\varphi_1(x_1)}{m_1}\cos\theta-\frac{\varphi_2(x_2+y)}{m_2}\cos\Theta+\frac{\varphi_1(x_1-y)}{m_1}\cos\theta_{\bp'\bk}\right].\label{eq:BE_6}
\end{eqnarray}
\end{subequations}
As in the main text, $\theta$ is the angle between the initial state momenta, $\bp$ and $\bk$, $\Theta$ is the scattering angle, $\phi$ is the angle between the planes formed by the initial and final momenta, respectively, $W(\theta,\phi)$ is the scattering probability, and $\cos\theta_{\bk'\bp}=\cos\theta_{\bp'\bk}=1+\cos\theta-\cos\Theta$, where $\theta_{{\bf m}{\bf n}}$ is the angle between vectors ${\bf n}$ and ${\bf m}$.  Furthermore,  $\mathcal{Q}$ represents the product of Fermi functions
\bea
	\mathcal{Q}(x_1,x_2,y)=n_F(x_1)n_F(x_2)\left[1-n_F(x_1-y)\right]\left[1-n_F(x_2+y)\right], \,n_F(x)=1/\left(e^x+1\right).
\eea
Equations~(\ref{eq:BE_5}) and (\ref{eq:BE_6}) can be further reduced to two coupled integral equations
 \begin{subequations}
 	\begin{eqnarray}
	1 &=&\tau^{-1}_{\rho 1} \frac{\pi^2+x_1^2}{2}\varphi_1(x_1)-\lambda_\rho\tau^{-1}_{\rho 1}\int du \, \mathcal{F}(x_1,u)\varphi_1(u)+(1-\lambda_\rho)\tau^{-1}_{\rho 2} \int du \,  \mathcal{F}(x_1,u) \varphi_2(u), \label{eq:Integral_rho_1}\\
	1 &=& \tau^{-1}_{\rho 2}\frac{\pi^2+x^2}{2}\varphi_2(x_2)-\lambda_\rho\tau^{-1}_{\rho 2} \int du \,  \mathcal{F}(x_2,u) \varphi_2(u)+(1-\lambda_\rho)\tau^{-1}_{\rho 1} \int du \,  \mathcal{F}(x_2,u) \varphi_1(u), \label{eq:Integral_rho_2}
\end{eqnarray}
 \end{subequations}
 where $\tau_{\rho 1,2}$ are given by Eq.~(\ref{tk1}) and its analog with $1\leftrightarrow 2$,  $\lambda_\rho$ is defined by Eq.~(\ref{lr}), and the kernel in the integrand reads
 \begin{subequations}
\bea
	 \mathcal{F}(x,u) &=& \frac{\cosh(x/2)}{\cosh(u/2)}\mathcal{G}(x-u),\label{kernel_F}\\
	 \mathcal{G}(x)&=&\frac{x}{2\sinh(x/2)}.\label{kernel_G}
\eea
\end{subequations}
To derive Eq.~(\ref{eq:Integral_rho_1}) and (\ref{eq:Integral_rho_2}),
%(\ref{eq:Integral_1quation_2}), 
we have used that
\begin{subequations}
	\begin{eqnarray}
	\int dx_2 \int dy \mathcal{Q}(x_1,x_2,y) \varphi_j(x_2) &=& -n_F(x_1)\left[1-n_F(x_1)\right] \int du \mathcal{F}(x_1,u) \varphi_j(u) \\
	\int dx_2 \int dy \mathcal{Q}(x_1,x_2,y) \varphi_j(x_1-y) &=& n_F(x_1)\left[1-n_F(x_1)\right] \int du \mathcal{F}(x_1,u) \varphi_j(u), \\
	\int dx_2 \int dy \mathcal{Q}(x_1,x_2,y) \varphi_j(x_2+y) &=& n_F(x_1) \left[1-n_F(x_1)\right]\int du \mathcal{F}(x_1,u) \varphi_j(u),
\end{eqnarray}
with $j=1,2$.
\end{subequations}
Defining $\Phi_j(x)=\varphi_j(x)/(\tau_{\rho, j}\cosh\frac{x}{2})$, we arrive at 
\begin{subequations}
	\begin{eqnarray}
	\frac{2}{\cosh(x_1/2)} &=& (\pi^2+x_1^2)\Phi_1(x_1)-2\int du \, \mathcal{G}(x_1-u)\left[\lambda_\rho\Phi_1(u)-(1-\lambda_\rho)\Phi_2(u)\right], \\
	\frac{2}{\cosh(x_2/2)} &=& (\pi^2+x_2^2)\Phi_2(x_2)-2\int du \, \mathcal{G}(x_2-u)\left[\lambda_\rho\Phi_2(u)-(1-\lambda_\rho)\Phi_1(u)\right].
\end{eqnarray}
\end{subequations}

The integral equations can be reduced to the differential ones for  Fourier transforms $\tilde{\Phi}_j(k)=\int dx\, \Phi_j(x)e^{ikx}$:\begin{subequations}
	\begin{eqnarray}
	\frac{d^2\tilde{\Phi}_1(k)}{dk^2}+\pi^2\left[\frac{2\lambda_\rho}{\cosh^2(\pi k)}-1\right]\tilde{\Phi}_1(k)-\pi^2\frac{2(1-\lambda_\rho)}{\cosh^2(\pi k)}\tilde{\Phi}_2(k) &=& -\frac{4\pi}{\cosh(\pi k)}, \\
	\frac{d^2\tilde{\Phi}_2(k)}{dk^2}+\pi^2\left[\frac{2\lambda_\rho}{\cosh^2(\pi k)}-1\right]\tilde{\Phi}_2(k)-\pi^2\frac{2(1-\lambda_\rho)}{\cosh^2(\pi k)}\tilde{\Phi}_1(k) &=& -\frac{4\pi}{\cosh(\pi k)}.
\end{eqnarray}
\end{subequations}
 Adding up the two equations above and introducing a variable $\zeta=\tanh(\pi k)$, we obtain an equation for 
 $\tilde{\Phi}(k)=\tilde{\Phi}_1(k)+\tilde{\Phi}_2(k)$
\begin{equation}
      \mathcal{L}\tilde{\Phi}(\zeta)+2(2\lambda_\rho-1)\tilde{\Phi}(\zeta)=-\frac{8}{\pi\sqrt{1-\zeta^2}},
\end{equation}
where the linear differential operator $\mathcal{L}$ is given by
\bea
\mathcal{L}=\frac{d}{d\zeta}\left[(1-\zeta^2)\frac{d}{d\zeta}\right]-\frac{1}{1-\zeta^2}.
\eea
The eigenfunctions of $\mathcal{L}$ are associated Legendre polynomials, $P^1_l(\zeta)$: 
$\mathcal{L}P^1_l(\zeta)=-l(l+1)P^1_l(\zeta)$, $l=1,2,3\dots$
Expanding $\tilde{\Phi}(\zeta)$ in series over $P^1_l(\zeta)$
\bea
	\tilde{\Phi}(\zeta) = \sum_{l} c_l P^1_l(\zeta),
\eea
and using  
the orthogonality relation
\bea
\int^1_{-1}d\zeta P^1_l(\zeta)P^1_m(\zeta)=\frac{2l(l+1)}{2l+1}\delta_{lm}\label{ortho}
\eea
and an identity
\bea
\int^1_{-1}d\zeta\frac{P^1_l(\zeta)}{\sqrt{1-\zeta^2}}=\Big \{
\begin{array}{cc}
	-2 ,& \mbox{if $l$ is odd}; \\
	0  ,& \mbox{if $l$ is even},
\end{array}
\eea
we obtain
%\begin{equation}
%\tilde{\Phi}(\zeta)=-\frac{8}{\pi}\sum_{n\in\text{odd}}\frac{2n+1}{n(n+1)}\,\frac{1}{n(n+1)-2(2\lambda_\rho-1)}\,P^1_n(\zeta),\label{Phiz}
%\end{equation}
%where $\lambda_\rho$ is defined by Eq.~(\ref{lr}) and we have used that
%\bea
%\int^1_{-1}d\zeta\frac{P^1_n(\zeta)}{\sqrt{1-\zeta^2}}=\Big \{
%\begin{array}{cc}
%	-2 ,& \mbox{if $n$ is odd}; \\
%	0  ,& \mbox{if $n$ is even}.
%\end{array}
%\eea
%Equation (\ref{Phiz}) coincides with 
Eq.~(\ref{phik}) of the main text. 

The electrical resistivity is given by
\begin{eqnarray}
	\rho &=& \frac{1}{ne^2}\left[\sum_{j=1,2}\frac{\tau_{\rho, j}}{m_j}\int dx \frac{\Phi_j(x)}{4\cosh(x/2)}\right]^{-1} 
	%&=&
	=\frac{1}{ne^2}\left[\sum_{j=1,2}\frac{\tau_{\rho, j}}{m_j}\int dk \frac{\tilde{\Phi}_j(k)}{4\cosh(\pi k)}\right]^{-1}\nn \\
	&=& \frac{1}{ne^2}\frac{m^2_1m^2_2T^2}{2\pi^4}\Big\langle \frac{W(\theta,\phi)}{\cos(\theta/2)}\Big\rangle \left[\int dk \frac{\tilde{\Phi}(k)}{\cosh(\pi k)}\right]^{-1} 
	%&=&
	= \frac{1}{ne^2}\frac{m^2_1m^2_2T^2}{2\pi^4}\Big\langle \frac{W(\theta,\phi)}{\cos(\theta/2)}\Big\rangle \left[\frac{1}{\pi}\int^1_{-1} d\zeta \frac{\tilde{\Phi}(\zeta)}{\sqrt{1-\zeta^2}}\right]^{-1}\nn \\
	&=& \frac{1}{ne^2}\frac{m^2_1m^2_2T^2}{8\pi^2} \Big\langle \frac{W(\theta,\phi)}{\cos(\theta/2)}\Big\rangle\left[\sum^\infty_{l=0}\frac{4l+3}{(l+1)(2l+1)}\,\frac{1}{(l+1)(2l+1)-(2\lambda_\rho-1)}\right]^{-1},
\end{eqnarray}
which reproduces the result in Eq.~(\ref{eq:rho_2}).

\subsection{Thermal conductivity}
In the same parametrization as in the previous section, Eqs.~(\ref{eq:BE_thermal_1}) and (\ref{eq:BE_thermal_2}) in the presence of a thermal gradient read
%To seek exact solution to thermal conductivity we start with
%\begin{subequations}
%	\begin{eqnarray}
%	-\frac{\xi_1}{T}\mathbf{v}_1\cdot\boldsymbol{\nabla}T\frac{\partial{n}}{\partial\xi_1} &=& -\mathcal{I}^{11}[g_1]-\mathcal{I}^{12}%[g_1,g_2], \\
%	-\frac{\xi_2}{T}\mathbf{v}_2\cdot\boldsymbol{\nabla}T\frac{\partial{n}}{\partial\xi_2} &=& -\mathcal{I}^{22}[g_2]-\mathcal{I}^{21}[g_1,g_2].
%\end{eqnarray}
%\end{subequations}
%The non-equilibrium part of the distribution function $g_i$ has the form 
%\begin{equation}
%	g_j=-\mathbf{v}_i\cdot\boldsymbol{\nabla}T\psi_i(\xi_i/T)/T,
%\end{equation}
%where $\psi$ is an odd function to be determined. We then have
\begin{subequations}
	\begin{eqnarray}
	\frac{1}{m_1}x_1\,n_F(x_1)\left[1-n_F(x_1)\right] &=& \frac{m^3_1T^2}{8\pi^4}\int \frac{d\Omega}{4\pi} \frac{W^{11}(\theta,\phi)}{\cos(\theta/2)} \int dx_2 \, dy \, \mathcal{Q}(x_1,x_2,y) \nonumber \\
	&& \times \frac{1}{m_1}\left[\psi_1(x_1)+\psi_1(x_2)\cos\theta-\psi_1(x_1-y)\cos\Theta-\psi_1(x_2+y)\cos\theta_{\bk'\bp}\right] \nonumber\\
	&&+ \frac{m_1m^2_2T^2}{8\pi^4}\int \frac{d\Omega}{4\pi} \frac{W(\theta,\phi)}{\cos(\theta/2)} \int dx_2 \, dy \, \mathcal{Q}(x_1,x_2,y) \nonumber\\
	&& \times \left[\frac{\psi_1(x_1)}{m_1}-\frac{\psi_2(x_2)}{m_2}\cos\theta-\frac{\psi_1(x_1-y)}{m_1}\cos\Theta+\frac{\psi_2(x_2+y)}{m_2}\cos\theta_{\bk'\bp}\right],\label{eq:BE_1} 
	\\ \nonumber\\
	\frac{1}{m_2}x_2\,n_F(x_2)\left[1-n_F(x_2)\right] &=& \frac{m^3_2T^2}{16\pi^4}\int \frac{d\Omega}{4\pi} \frac{W^{22}(\theta,\phi)}{\cos(\theta/2)} \int dx_1 \, dy \, \mathcal{Q}(x_1,x_2,y) \nonumber \\
	&& \times \frac{1}{m_2}\left[\psi_2(x_2)+\psi_2(x_1)\cos\theta-\psi_2(x_2+y)\cos\Theta-\psi_2(x_1-y)\cos\theta_{\bp'\bk}\right] \nonumber\\
	&&+ \frac{m_2m^2_1T^2}{16\pi^4}\int \frac{d\Omega}{4\pi} \frac{W(\theta,\phi)}{\cos(\theta/2)} \int dx_1 \, dy \, \mathcal{Q}(x_1,x_2,y) \nonumber\\
	&& \times \left[\frac{\psi_2(x_2)}{m_2}-\frac{\psi_1(x_1)}{m_1}\cos\theta-\frac{\psi_2(x_2+y)}{m_2}\cos\Theta+\frac{\psi_1(x_1-y)}{m_1}\cos\theta_{\bp'\bk}\right].\label{eq:BE_2}
\end{eqnarray}
\end{subequations}
Eqs.~(\ref{eq:BE_1}) and (\ref{eq:BE_2}) can be further reduced to two coupled integral equations,
 \begin{subequations}
 	\begin{eqnarray}
	\tau_{\kappa 1} \, x_1 &=& \frac{\pi^2+x_1^2}{2}\psi_1(x_1)-\lambda_{\kappa 1}\int du \, \mathcal{F}(x_1,u)\psi_1(u)+\beta_{\kappa 1} \int du \,  \mathcal{F}(x_1,u) \psi_2(u), \label{eq:Integral_1quation_1}\\
	\tau_{\kappa 2} \, x_2 &=& \frac{\pi^2+x^2}{2}\psi_2(x_2)-\lambda_{\kappa 2}\int du \,  \mathcal{F}(x_2,u) \psi_2(u)+\beta_{\kappa 2} \int du \,  \mathcal{F}(x_2,u) \psi_1(u), \label{eq:Integral_1quation_2}
\end{eqnarray}
 \end{subequations}
where parameters $\tau_{\kappa 1,2}$, $\lambda_{\kappa 1,2}$, and $\beta_{\kappa 1,2}$ are given by Eqs.~(\ref{tk2}), (\ref{lk1}) and (\ref{bk1}), respectively, and their analogs with $1\leftrightarrow 2$, and kernels $\mathcal{F}(x,u)$ and $\mathcal{G}(x)$ are defined by Eqs.~(\ref{kernel_F}) and (\ref{kernel_G}), respectively.
%the kernel in the integrand reads
%\bea
%	 \mathcal{F}(x,u) = \frac{\cosh(x/2)}{\cosh(u/2)}\mathcal{G}(x-u),\,\mathcal{G}(x)=\frac{x}{2\sinh(x/2)},
%\eea
%and we have used the fact that
%\begin{subequations}
%	\begin{eqnarray}
%	\int dx_2 \int dy \mathcal{Q}(x_1,x_2,y) \psi_i(x_2) &=& -n_F(x_1)\left[1-n_F(x_1)\right] \int du \mathcal{F}(x_1,u) \psi_i(u) \\
%	\int dx_2 \int dy \mathcal{Q}(x_1,x_2,y) \psi_i(x_1-y) &=& n_F(x_1)\left[1-n_F(x_1)\right] \int du \mathcal{F}(x_1,u) \psi_i(u), \\
%	\int dx_2 \int dy \mathcal{Q}(x_1,x_2,y) \psi_i(x_2+y) &=& n_F(x_1) \left[1-n_F(x_1)\right]\int du \mathcal{F}(x_1,u) \psi_i(u).
%\end{eqnarray}
%\end{subequations}
Defining $\Psi_j(x)=\psi_j(x)/(\tau_{\kappa,j}\cosh\frac{x}{2})$, we arrive at
\begin{subequations}
	\begin{eqnarray}
	\frac{2x_1}{\cosh(x_1/2)} &=& (\pi^2+x_1^2)\Psi_1(x_1)-2\int du \, \mathcal{G}(x_1-u)\left[\lambda_{\kappa 1}\Psi_1(u)-\frac{\beta_{\kappa 1}\tau_{\kappa 2}}{\tau_{\kappa 1}}\Psi_2(u)\right], \label{eq:Integral_1quation_3}\\
	\frac{2x_2}{\cosh(x_2/2)} &=& (\pi^2+x_2^2)\Psi_2(x_2)-2\int du \, \mathcal{G}(x_2-u)\left[\lambda_{\kappa 2}\Psi_2(u)-\frac{\beta_{\kappa 2}\tau_{\kappa 1}}{\tau_{\kappa 2}}\Psi_1(u)\right]. \label{eq:Integral_1quation_4}
\end{eqnarray}
\end{subequations}
After a  Fourier transformation, $\tilde{\Psi}_j(k)=\int dx \, \Psi_j(x)e^{ikx}$, Eqs.~(\ref{eq:Integral_1quation_3}) and (\ref{eq:Integral_1quation_4}) become
%, and changing the variable $\zeta=\tanh(\pi k), \,-1<\zeta<1$, we then have
\begin{subequations}
\begin{eqnarray}
    \mathcal{L}\tilde{\Psi}_1(\zeta)+2\lambda_{\kappa 1}\tilde{\Psi}_1(\zeta)-2\frac{\beta_{\kappa 1}\tau_{\kappa 2}}{\tau_{\kappa 1}} \tilde{\Psi}_2(\zeta) &=& -i\frac{4\zeta}{\sqrt{1-\zeta^2}}, \\
    \mathcal{L}\tilde{\Psi}_2(\zeta)+2\lambda_{\kappa 2}\tilde{\Psi}_2(\zeta)-2\frac{\beta_{\kappa 2}\tau_{\kappa 1}}{\tau_{\kappa 2}}\tilde{\Psi}_1(\zeta)&=& -i\frac{4\zeta}{\sqrt{1-\zeta^2}}.
\end{eqnarray}
\end{subequations}
Expanding $\tilde{\Psi}_{\kappa 1}(\zeta)$ and $\tilde{\Psi}_{\kappa 2}(\zeta)$ in series of 
%with 
the associated Legendre polynomial $P^1_l(\zeta)$, 
\bea
	\tilde{\Psi}_1(\zeta) = i \sum_{n} a_l P^1_l(\zeta),\,\tilde{\Psi}_2 (\zeta) = i \sum_{l} b_l P^1_l(\zeta),\label{series}
\eea
we then obtain
\begin{subequations}
	\begin{eqnarray}
	\left[l(l+1)-2\lambda_{\kappa 1}\right]a_l+\frac{2\beta_{\kappa 1}\tau_{\kappa 2}}{\tau_{\kappa 1}}b_l &=& \frac{2(2l+1)}{l(l+1)}\int^1_{-1}d\zeta \frac{\zeta P^1_l(\zeta)}{\sqrt{1-\zeta^2}}, \\
    \frac{2\beta_{\kappa 2}\tau_{\kappa 1}}{\tau_{\kappa 2}} a_l+\left[l(l+1)-2\lambda_{\kappa 2}\right]b_l&=& \frac{2(2l+1)}{l(l+1)} \int^1_{-1}d\zeta \frac{\zeta P^1_l(\zeta)}{\sqrt{1-\zeta^2}}.
\end{eqnarray}
\end{subequations}
Solving for $a_l$ and $b_l$, we find
\begin{subequations}
	\begin{eqnarray}
	a_l &=& -\frac{l(l+1)-2\lambda_{\kappa 2}-2\beta_{\kappa 1}\tau_{\kappa 2}/\tau_{\kappa 1}}{\left[l(l+1)-2\lambda_{\kappa 1}\right]\left[l(l+1)-2\lambda_{\kappa 2}\right]-4\beta_{\kappa 1}\beta_{\kappa 2}} \, \frac{4(2l+1)}{l(l+1)}, \\
	b_l &=& -\frac{l(l+1)-2\lambda_{\kappa 1}-2\beta_{\kappa 2}\tau_{\kappa 1}/\tau_{\kappa 2}}{\left[l(l+1)-2\lambda_{\kappa 1}\right]\left[l(l+1)-2\lambda_{\kappa 2}\right]-4\beta_{\kappa 1}\beta_{\kappa 2}} \, \frac{4(2l+1)}{l(l+1)},
\end{eqnarray}
\end{subequations}
where we have again used Eq.~(\ref{ortho}) and another identity
\bea
\int^1_{-1}d\zeta\frac{\zeta P^1_l(\zeta)}{\sqrt{1-\zeta^2}}=\Big \{
\begin{array}{cc}
	-2 ,& \mbox{if $l$ is even}; \\
	0  ,& \mbox{if $l$ is odd}.
\end{array}
\eea
Substituting $a_l$ and $b_l$ into Eq.~(\ref{series}), we reproduce Eqs.~(\ref{psi1}) and (\ref{psi2}) of the main text.
%  Finally, we obtain
%\begin{subequations}
%	\begin{eqnarray}
%	    \tilde{\Psi}_1(\zeta) &=& -4i \sum_{n, even} \frac{n(n+1)-2\lambda_{\kappa 2}-2\beta_{\kappa 1}\tau_{\kappa 2}/\tau_{\kappa 1}}%{\left[n(n+1)-2\lambda_{\kappa 1}\right]\left[n(n+1)-2\lambda_{\kappa 2}\right]-4\beta_{\kappa 1}\beta_{\kappa 2}} \, \frac{2n+1}{n(n+1)} P^1_n(\zeta),\\
%	     \tilde{\Psi}_2(\zeta) &=&  -4i \sum_{n,even} \frac{n(n+1)-2\lambda_{\kappa 1}-2\beta_{\kappa 2}\tau_{\kappa 1}/\tau_{\kappa 2}}{\left[n(n+1)-2\lambda_{\kappa 1}\right]\left[n(n+1)-2\lambda_{\kappa 2}\right]-4\beta_{\kappa 1}\beta_{\kappa 2}} \, \frac{2n+1}{n(n+1)} P^1_n(\zeta).
 %   \end{eqnarray}
%\end{subequations}

The thermal conductivity is found as 
\begin{eqnarray}
	\kappa &=& nT\left[\sum_{j=1,2}\frac{\tau_{\kappa, j}}{m_j}\int dx \frac{x}{4\cosh(x/2)}\Psi_j(x)\right]
	%, \\
	%&=& 
	=-i\pi nT\left[\sum_{j=1,2}\frac{\tau_{\kappa, j}}{m_j}\int dk \frac{\sinh(\pi k)}{4\cosh^2(\pi k)}\tilde{\Psi}_j(k)\right]\nn \\
	&=& -i\frac{nT}{4}\left[\sum_{j=1,2}\frac{\tau_{\kappa, j}}{m_j}\int^1_{-1}d\zeta\frac{\zeta\tilde{\Psi}_j(\zeta)}{\sqrt{1-\zeta^2}}\right] \nn\\
	&=& \frac{4\pi^4n}{m^2_1T}\frac{1}{m^2_1\Big\langle W^{11}(\theta,\phi)/\cos\frac{\theta}{2}\Big\rangle+m^2_2\Big\langle W(\theta,\phi)/\cos\frac{\theta}{2}\Big\rangle}\times \nonumber \\
	&& \sum^\infty_{l=0} \frac{4l+5}{(l+1)(2l+3)}\frac{(l+1)(2l+3)-\lambda_{\kappa 2}-\beta_{\kappa 1}\tau_{\kappa 2}/\tau_{\kappa 1}}{\left[(l+1)(2l+3)-\lambda_{\kappa 1}\right]\left[(l+1)(2l+3)-\lambda_{\kappa 2}\right]-\beta_{\kappa 1}\beta_{\kappa 2}}\nonumber \\
	&& +\frac{4\pi^4n}{m^2_2T}\frac{1}{m^2_2\Big\langle W^{22}(\theta,\phi)/\cos\frac{\theta}{2}\Big\rangle+m^2_1\Big\langle W(\theta,\phi)/\cos\frac{\theta}{2}\Big\rangle} \times \nonumber \\
	&&  \sum^\infty_{l=0} \frac{4l+5}{(l+1)(2l+3)}\frac{(l+1)(2l+3)-\lambda_{\kappa 1}-\beta_{\kappa 2}\tau_{\kappa 1}/\tau_{\kappa 2}}{\left[(l+1)(2l+3)-\lambda_{\kappa 1}\right]\left[(l+1)(2l+3)-\lambda_{\kappa 2}\right]-\beta_{\kappa 1}\beta_{\kappa 2}},
\end{eqnarray}
which reproduces Eq.~(\ref{eq:kappa_2}) in the main text.
\ewt

%\bibliography{/Users/maslov/Documents/Bibliography/dm_references}
%\bibliography{/Users/sclee/Documents/Bibliography/dm_references}
\bibliography{dm_references}

\end{document}